%% file: main.tex
\documentclass[lettersize,journal]{IEEEtran}

\input{packages.tex}

\input{macros.tex}

\begin{document}

\newcommand{\nmctitle}{Scalable and RISC-V Programmable Near-Memory Computing Architectures for Edge Nodes}
\title{\nmctitle}

\author{
    Michele Caon\orcidlink{0009-0009-4446-8389}, Clément Choné\orcidlink{0009-0000-2625-9019}, Pasquale Davide Schiavone\orcidlink{0000-0003-2931-0435}, Alexandre Levisse\orcidlink{0000-0002-8984-9793}, Guido Masera\orcidlink{0000-0003-2238-9443}~\IEEEmembership{Senior member, IEEE}, Maurizio Martina\orcidlink{0000-0002-3069-0319}~\IEEEmembership{Senior member, IEEE}, David Atienza\orcidlink{0000-0001-9536-4947}~\IEEEmembership{Fellow, IEEE}%

    \thanks{M. Caon is with the Embedded Systems Laboratory (ESL), EPFL, 1015 Lausanne, Switzerland and with the Very Large Scale Integration Laboratory (VLSI Lab), DET, Politecnico di Torino, 10129 Torino, Italy (e-mail: michele.caon@epfl.ch). C. Choné, P. D. Schiavone, A. Levisse and D. Atienza are with the Embedded Systems Laboratory (ESL), EPFL (e-mail: clement.chone@epfl.ch; davide.schiavone@epfl.ch; alexandre.levisse@epfl.ch; david.atienza@epfl.ch). G. Masera and M. Martina are with the Very Large Scale Integration Laboratory (VLSI Lab), DET, Politecnico di Torino (e-mail: guido.masera@polito.it; maurizio.martina@polito.it).}%

    \thanks{\textit{M. Caon and C. Choné contributed equally to this work.}}
}


\maketitle

\begin{abstract} 
    The widespread adoption of data-centric algorithms, particularly \acf*{ai} and \acf*{ml}, has exposed the limitations of centralized processing infrastructures, driving a shift towards edge computing. This necessitates stringent constraints on energy efficiency, which traditional von Neumann architectures struggle to meet. The \acf*{cim} paradigm has emerged as a better candidate due to its efficient exploitation of the available memory bandwidth. However, existing \acs*{cim} solutions require a high implementation effort and lack flexibility from a software integration standpoint. This work proposes a novel, software-friendly, general-purpose, and low-integration-effort \acf*{nmc} approach, paving the way for the adoption of \acs*{cim}-based systems in the next generation of edge computing nodes. Two architectural variants, \caesar and \carus, are proposed and characterized to target different trade-offs in area efficiency, performance, and flexibility, covering a wide range of embedded microcontrollers.
    Post-layout simulations show up to \SI{28.0}{\times} and \SI{53.9}{\times} lower execution time and \SI{25.0}{\times} and \SI{35.6}{\times} higher energy efficiency at system level, respectively, compared to the execution of the same tasks on a state-of-the-art \riscv \acs*{cpu} (RV32IMC).
    \carus achieves a peak energy efficiency of \SI{306.7}{\giga OPS \per\watt} in 8-bit matrix multiplications, surpassing recent state-of-the-art in- and near-memory circuits.
\end{abstract}

\begin{IEEEkeywords}
    Near-Memory Computing, Edge Computing, \riscv, Embedded Systems, Microcontrollers.
\end{IEEEkeywords}

\input{introduction.tex}

\input{related-works.tex}

\input{architectures.tex}
\input{implementation.tex}

\input{performance.tex}
\input{conclusion.tex}


\input{acronyms.tex}

\bibliographystyle{IEEEtran}
\bibliography{bibliography.bib}

\input{authors.tex}

\vfill

\end{document}

%% file: packages.tex
\usepackage[utf8]{inputenc}
\usepackage[T1]{fontenc}
\usepackage[english]{babel}

\usepackage{amsmath,amsfonts}
\usepackage{array}
\usepackage{textcomp}
\usepackage{stfloats}
\usepackage{url}
\usepackage{verbatim}
\usepackage{graphicx}
\usepackage{cite}

\usepackage[frozencache=true,cachedir=minted-cache]{minted}
\usepackage{algorithm}
\usepackage[noend]{algpseudocode}

\usepackage{booktabs}
\usepackage{tabularx}
\usepackage{multirow}
\usepackage{multicol}
\usepackage{colortbl}
\usepackage[para]{threeparttable}
\usepackage{changepage}
\usepackage{soul}
\usepackage{adjustbox}
\usepackage{float}

\usepackage[table]{xcolor}
\usepackage{cellspace}
\usepackage{svg}
\usepackage{orcidlink}
\usepackage{placeins}
\usepackage{tikz}
\usepackage{amssymb}
\usepackage[font=small,labelfont=bf]{caption}
\usepackage{subcaption}

\usepackage[capitalise]{cleveref}

\usepackage{siunitx}
\usepackage{xfrac}
\usepackage{xspace}
\usepackage[inline]{enumitem}
\usepackage{footnote}
\usepackage{textgreek}
\usepackage[nolist]{acronym}
\usepackage{lipsum}

\def\BibTeX{{\rm B\kern-.05em{\sc i\kern-.025em b}\kern-.08em
    T\kern-.1667em\lower.7ex\hbox{E}\kern-.125emX}}

\algrenewcommand\alglinenumber[1]{\scriptsize #1:}

\algrenewcommand\algorithmicloop{\textbf{loop}\xspace}
\algnewcommand{\algorithicgoto}{\textbf{goto}}%
\algnewcommand{\Goto}[1]{\algorithicgoto~\ref{#1}}%
\algnewcommand{\algorithmicnot}{\textbf{not}}
\algdef{SE}[IF]{IfNot}{EndIf}[1]{\algorithmicif\ \algorithmicnot\ #1\ \algorithmicthen}{\algorithmicend\ \algorithmicif}%
\makeatletter
\ifthenelse{\equal{\ALG@noend}{t}}%
  {\algtext*{EndIf}}
  {}%
\makeatother

\newcommand{\tnotemid}[1]{\textsuperscript{\TPTtagStyle{#1}}} 
\newcommand{\tabitem}{~\llap{\textbullet}~} 

\definecolor{lightgray}{gray}{0.98}
\definecolor{cpu}{HTML}{434343}
\definecolor{carus}{HTML}{007480}
\definecolor{caesar}{HTML}{6d1a36}

%% file: macros.tex
\newcommand{\alu}{\ac{alu}\xspace}
\newcommand{\caesar}{NM-Caesar\xspace}
\newcommand{\carus}{NM-Carus\xspace}
\newcommand{\cpu}{\ac{cpu}\xspace}
\newcommand{\ecpu}{\ac{ecpu}\xspace}
\newcommand{\emem}{\ac{emem}\xspace}

\newcommand{\heeperator}{HEEPerator\xspace}
\newcommand{\imc}{\ac{imc}\xspace}
\newcommand{\isa}{\ac{isa}\xspace}
\newcommand{\iot}{\ac{iot}\xspace}
\newcommand{\mcu}{\ac{mcu}\xspace}
\newcommand{\nmc}{\ac{nmc}\xspace}
\newcommand{\riscv}{RISC\babelhyphen{nobreak}V\xspace}
\newcommand{\rvv}{\ac{rvv}\xspace}

\newcommand{\soc}{\ac{soc}\xspace}
\newcommand{\sram}{\ac{sram}\xspace}
\newcommand{\vpu}{\ac{vpu}\xspace}
\newcommand{\vrf}{\ac{vrf}\xspace}
\newcommand{\xisa}{\texttt{xvnmc}\xspace}



%% file: introduction.tex
\section{Introduction}
\label{sec:intro}

\IEEEPARstart{T}{he} last decades have witnessed a shift in the computing paradigm towards data-driven workloads, such as \acp{ann}. These computationally expensive algorithms offer advanced functionalities in a wide range of applications by exploiting an exponentially increasing volume of data from the internet and \iot devices.
In this context, edge computing has gained traction, driving the exploration of innovative computing architectures that provide superior energy efficiency and performance compared to traditional systems. These advances are crucial in overcoming the throughput, latency, energy, power, and privacy limitations of centralized computing infrastructures while meeting the increased computational demands for real-time data processing in fields such as biosignals\cite{sussillo2024generic}, audio\cite{VAD}, video, etc.
The accessible manufacture and deployment of low-power, high-performance, and software-friendly edge computing devices for the new era of \ac{ai} is the main motivation behind the research presented in this paper.
The traditional von Neumann architecture, which has been the backbone of computing systems for decades, is inherently inefficient for data-intensive workloads due to the constant need to move data and instructions between the system memory hierarchy and the \cpu registers\cite{VN_bottleneck, computing_energy_problem}. This is exacerbated by \ac{sram} integration technologies not keeping pace with logic scaling, an issue known as the \emph{memory wall}\cite{memory_wall}. As a consequence, \ac{sram} accesses typically account for $100\times$ the energy of the arithmetic operations in the \cpu, as noted by J. Hennessy and D. Patterson\cite{hennessy_computer_2017}.
The \emph{\ac{cim}} paradigm has been proposed as a solution to this problem in the form of \acf{imc} and \acf{nmc}\cite{survey_imc_nmc,survey_nmc, 
data_mvt_expensive}.
Their common intuition is to move processing closer to the data to alleviate the instruction fetch overhead and the pressure on the system bus significantly while optimizing the utilization of the available internal memory bandwidth. 

The preserved versatility of flexible memory access in near-memory systems suits the semantics of traditional programmable systems, where data processing is expressed through instructions operating on data from various memory locations. Consequently, the optimizations available in traditional von Neumann architectures, such as \ac{simd} and vector operations, can be easily applied to \ac{cim} systems. This was demonstrated in \cite{wang2024vecim}, where the energy efficiency of a vector coprocessor based on Ara \cite{cavalcante2020ara} was further enhanced by integrating parts of the arithmetic circuitry within the \vrf memory banks.

Controlling operations on \nmc \acp{ic} remains a challenge. The simplest control strategy involves streaming individual micro-operations to the \nmc unit using an external \cpu or \ac{dma} controller. This approach is very area-efficient, does not rely on a dedicated external controller, and scales well for regular packed-\acs{simd} applications. However, complex runtime control requires either the \cpu to spend significant time encoding such operations at runtime, decreasing system efficiency, or relying on predefined command sequences, resulting in substantial code size overhead. In both cases, additional instructions and data must be fetched from the main memory before sending commands to the \nmc \acp{ic}, reducing energy efficiency. Alternatively, custom controllers can assemble the necessary commands in hardware, increasing energy efficiency and throughput at the cost of reduced flexibility.
To avoid microcontrolling the \nmc unit, more complex and versatile controllers can be employed. For example, a dedicated \cpu can be coupled with a \nmc block to run small programs via a custom controller memory interface, avoiding standard load/store memory-mapped operations. This enables the execution of more complex programs independently from the host system, similar to a conventional accelerator, with the advantage of processing private data only
 with closer processing elements, resulting in fewer data movements.
However, such controllers trade flexibility for increased area and power, making them unsuitable for ultra-low-power, edge-oriented \acp{soc}.

In this work, we elaborate on this concept to present two architectural variations of a novel \nmc approach and their corresponding \ac{ic} implementations, which target different requirements at the system and application level, such as complexity, performance, and integration with existing platforms:

\begin{itemize}
     \item \emph{\caesar}, an area-efficient, \ac{simd}-enabled \nmc unit micro-controlled by the host system, targeting regular TinyML\footnote{TinyML applications here refer to compact machine-learning algorithms deployed on energy- and area-constrained edge devices.} benchmarks (i.e., \acs{ai}-based biomedical application kernels with a regular control flow) such as min/max search algorithms for peak detection\cite{Heartbeat}, and lightweight \acp{ann}\cite{Litenet} used in arrhythmia detection.
     \item \emph{\carus}, a fully-autonomous, vector-capable \riscv-based programmable \nmc unit targeting highly-parallel and complex TinyML applications \cite{TinyML} across domains involving performance-critical and computationally-intensive workloads (i.e., \acp{dnn} \cite{SeizureDetCNN, acharya2017automated} or tasks with a data-dependent control flow).
\end{itemize}

Both \carus and \caesar are designed to be energy-efficient computing solutions as close as possible to a drop-in alternative to a traditional embedded \sram bank both from a physical and functional standpoint. They offer an \acs{sram}-compatible interface to the host system and have a functionally transparent \textit{memory} operating mode alongside their \textit{computing} mode. Their focus on programmability and ease of integration represents a significant advancement to address the challenges that prevent available \ac{cim} architectures from achieving broader adoption, as highlighted in \cite{khan2024landscape}.

In particular, the main contributions of the presented work on emerging computing paradigms are as follows:
\begin{itemize}
    \item A software-friendly, low-cost, and low-integration-effort approach to \nmc in the context of general-purpose, low-power edge devices. Its effectiveness is demonstrated through the implementation of two architectural variations targetting various classes of embedded \acp{soc}.
    \item Conception of a code-efficient \riscv custom \isa extension to support a flexible set of vector operations on programmable \ac{cim} architectures. The proposed extension provides a vector-based view of the host's compute memory without resulting in the overhead of explicit vector load/store operations.
    \item An in-depth, quantitative, and vertical analysis of the impact and benefits of replacing conventional \ac{sram} banks with the proposed \nmc macros in low-power \acp{mcu}.
\end{itemize}

The remainder of this paper is organized as follows. \Cref{sec:related} reviews existing works on improving the performance and energy efficiency of edge-computing systems, focusing on \ac{cim} devices. \Cref{sec:arch,sec:impl} detail the architecture and physical implementation of \caesar and \carus. \Cref{sec:perf} describes the experimental setup, benchmarking methodology, and results. \Cref{sec:conclusion} summarizes the main conclusions for the \nmc paradigm emerging from this work.

%% file: related-works.tex
\section{Related Works}
\label{sec:related}

\noindent To overcome the efficiency limitations of traditional von Neumann systems in handling data-centric workloads, recent research has explored alternative and more efficient computing paradigms. Among these, application- or domain-specific accelerators and \ac{cim} devices are the most prominent ones, and the closest to the proposed \nmc architecture presented in this work.

\subsection{Energy-efficient Edge Accelerators}
\noindent Fixed-function \acp{asic} have been proposed for maximum performance and energy efficiency, whose predominant examples are \ac{ann}\cite{cnn-accel,cnn-sa,cnn-sa-2} accelerators, specialized \acp{dsp} for real-time data elaboration kernels such as\acp{fft}\cite{fft}, or cryptographic coprocessors\cite{keccak}. These \acp{ic} leverage highly parallel computing engines and optimized memory management units that maximize data reuse to achieve superior performance and energy efficiency, at the cost of reduced flexibility.

For scenarios requiring greater versatility, reconfigurable solutions such as embedded \acp{fpga}\cite{efpga} and \acp{cgra}\cite{cgra-lab,cgra} can handle a wider range of workloads, although with a higher area and power overhead compared to \acp{asic}. From a functional perspective, they often suffer from suboptimal resource utilization due to the complexity of the application mapping process, especially in on-line-reconfigurable \acp{cgra} where spatial and temporal mapping constraints coexist.
More conventional architectures preserve programmability while overcoming the instruction fetch overhead by offloading tasks to on-chip domain-specific processors\cite{dsp_coproc}, \acf{vpu}\cite{vicuna,cavalcante2020ara}, or multi-core clusters with optimized data interconnects\cite{pulp}. These subsystems efficiently execute critical parts of the workload, reducing the pressure on the system bus and the memory hierarchy, thereby improving overall performance and energy efficiency.
However, they still rely on a dedicated local memory that requires specific management in the application software, resulting in additional instructions and data movement. To address these limitations and leverage the benefits of domain-specific architectures while significantly reducing data movement, our work aims to maintain the flexibility and ease of implementation of programmable accelerators while exploiting the advantages of the aforementioned \ac{cim} paradigm. With this objective, it implements an innovative integration model where compute-capable drop-in replacements for conventional \acp{sram} can be arbitrarily programmed to process data in place, with minimal impact on the host's software stack and physical characteristics.

\subsection{Compute Memories}
\noindent \imc and \nmc circuits enable concurrent data access from several memory banks and, in the case of \imc devices, also exploit the internal parallelism and structure of the memory arrays to enhance arithmetic throughput with lower area overhead compared to conventional memory-mapped accelerators. Over the years, this concept has been applied at various levels of the memory hierarchy.
\emph{\ac{envm}} technologies such as \ac{mram}\cite{MRAM}, \ac{pcm}\cite{PCM}, and \ac{rram}\cite{RRAM} have emerged as next-generation memory solutions enabling fast, power-efficient \ac{mac} operations in the analog domain, making them suitable for applications tolerating inexact results, such as \acp{ann}. However, their stochastic nature also leads to reliability, uniformity, and portability issues, making them unsuitable for memory-critical applications\cite{eNVM}. 
Consequently, research has also focused on \acs*{dram}- and \acs*{sram}-based solutions\cite{DRAM_berkley,CSRAM}.
\emph{\acfp{dram}} offer high density and large storage capacity, reducing data transfers from other memory devices, which is beneficial for data-intensive \ac{cim} scenarios. However, their high memory access energy limits their use in energy-constrained TinyML systems.

Conversely, \emph{\acfp{sram}} provide lower-energy and faster memory accesses, motivating studies on \acs{sram}-based \imc or \nmc architectures\cite{blade-tc}. As a result, \acs{sram} has become the predominant technology for data-intensive applications\cite{khan2024landscape}. In \imc devices, the integration of arithmetic circuitry directly within the memory array allows fine-grained optimization, producing superior performance and energy efficiency\cite{khoram2017challenges}. However, this approach is highly invasive at the system level, introducing significant portability challenges and imposing strict data placement constraints that degrade application-level performance, as demonstrated in \cref{sec:soa-comparison}.
In contrast, \nmc architectures place digital processing units outside the memory periphery. This choice leverages a standard digital integration flow, reducing implementation effort and enhancing portability across technology nodes. These advantages motivated the adoption of an \nmc approach for the architectures proposed in \cref{sec:arch}.
Hybrid \acs{imc}/\ac{nmc} solutions\cite{wang2024vecim,CSRAM,survey_imc_nmc,gauchi2020reconfigurable} have been explored as alternatives.
In \cite{wang2024vecim}, Wang et al. propose Vecim, a \riscv vector coprocessor leveraging a \acs{cim}-based \vrf. Its modular design features configurable vector lanes that implement in-memory \ac{mac} operations through bit-line computing. This design achieves a state-of-the-art peak energy efficiency of \SI{289.1}{\giga OPS \per\watt} in \SI{65}{\nano\meter} CMOS technology during an 8-bit matrix multiplication. In comparison, the proposed \carus macro achieves \SI{306.7}{\giga OPS \per\watt} under similar conditions to a full \nmc design giving priority to memory capacity and reduced data movement over high parallelization of processing elements, as explained in \cref{sec:carus-arch}.
Kooli et al.\cite{CSRAM} present a Computational \sram (C-SRAM) solution combining \imc and \nmc to perform vector operations on memory rows. However, this approach suffers from low bitcell density due to data replication and the use of larger two-port memories.
Gauchi et al. \cite{gauchi2020reconfigurable} propose a workload-aware, reconfigurable multi-tile C-SRAM architecture to enhance vectorization, albeit with higher area and power overhead, reduced versatility, and suboptimal utilization of computing resources on small kernels. C-SRAM acts as a memory-mapped accelerator that is micro-controlled with commands encoded in bus write transactions. \caesar uses a similar control strategy while achieving higher bitcell density and energy efficiency thanks to smaller single port memories at the cost of no support for vector operations and lower throughput, as reported in \cref{sec:soa-comparison}.

From a technological standpoint, these hybrid approaches mitigate many challenges associated with fully \imc-based solutions. By placing most of the processing hardware outside the memory arrays, they maintain a clear separation between physical memory and arithmetic circuitry. This separation simplifies validation and verification while enabling faster architectural exploration. However, they still present integration challenges at the system level, which are addressed in this work.
As discussed in \cite{khan2024landscape}, strict data placement constraints, and the need for dedicated control schemes significantly hinder the development of a robust software deployment ecosystem. These factors, rather than technological limitations alone, remain a major barrier to the widespread adoption of \ac{cim} devices in real-world applications.
The \ac{cim} paradigm introduced in this paper aims to overcome these limitations by enabling transparent integration of compute-memory elements within the memory subsystem of the host \ac{soc} (\cref{sec:arch}).
The proposed \nmc architectures provide a low-impact integration effort from both software and hardware perspectives (\cref{sec:impl}), combining the efficiency of \acs{sram}-based \ac{cim} with the flexibility and portability of programmable accelerators.

%% file: architectures.tex
\section{\acl*{nmc} Architectures}
\label{sec:arch}

\begin{figure}[htb]
    \centering
    \includesvg[width=.95\linewidth]{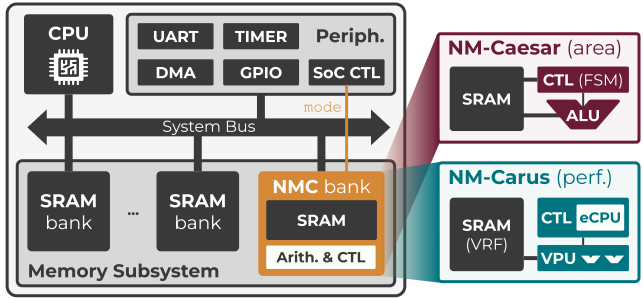}
    \caption{Top-level block diagram of a \acs{nmc}-enhanced \acs{mcu} hosting \caesar (area-critical implementations) or \carus (performance-oriented applications) as part of its memory subsystem.}
    \label{fig:nmc-system-top}
\end{figure}

\noindent The main goal of the proposed \nmc paradigm is to bring the efficiency of \ac{cim} to existing \acp{soc} with minimal integration effort, providing an easy-to-deploy solution to the growing computational needs of next-generation edge computing devices. To achieve this, the \nmc device must adhere to two fundamental requirements:
\begin{enumerate*}[label=(\arabic*)]
    \item functionally, it is part of the host system's memory space and should operate like a conventional memory during normal use;\label{item:functional}
    \item from a physical implementation perspective, it represents a direct replacement for a standard \sram memory bank, with similar area and timing characteristics.\label{item:physical}
\end{enumerate*}
An overview of the resulting integration model is shown in \cref{fig:nmc-system-top}.

For straightforward functional integration \labelcref{item:functional}, the proposed \nmc devices connect to the system bus with a slave interface, similar to conventional memories. A transparent \emph{memory} mode provides read and write access to the memory content, while a \emph{computing} or \emph{configuration} mode enables streaming or programming of arbitrary processing kernels. The data is processed in place within the \ac{ic} private memory, which is managed and populated by the host \cpu or \ac{dma} engine. The computation results are directly accessible by the host system, eliminating additional data movement.
All interactions take place on a memory-like interface, whose data write bus is used to encode and stream instructions when in \emph{computing} mode. An additional \texttt{imc} pin, routed to a dedicated configuration register in the host system, switches the operating mode and is controlled by software, without altering the bus protocol. Unlike other implementations where the address bus encodes instructions, this solution avoids fragmentation of the address space and consequent mapping constraints when linking the firmware. The cost of setting and resetting the \texttt{imc} configuration register is negligible in common edge-computing workloads, where a long burst of operations is executed after writing and before reading back the memory entries.

The physical integration \labelcref{item:physical} of the proposed \nmc \acp{ic} is facilitated by the standard digital implementation flow for logic and the use of foundry-provided \sram compilers for internal memories. This improves technological portability compared to \imc solutions that rely on custom memory arrays. The control and data processing units are carefully designed to maintain the timing characteristics of a reference conventional \sram memory with the same capacity, while also minimizing the area overhead. Particular attention is paid to preserving the reference input and output delays, ensuring compatibility when replacing a standard memory bank with the \nmc device.

Having established the system integration model and its implication in functional and physical aspects of the design, it is possible to motivate the design choices behind the proposed \nmc architectures. The chosen fully digital \nmc paradigm enables the development of configurable and scalable architectures, possibly covering diverse performance requirements. However, the physical constraints of the smallest class of \iot devices conflict with the level of flexibility that brings the most advantages to larger, performance-oriented systems. To address this trade-off, two distinct \nmc architectures were designed and implemented. \caesar targets low cost \acp{soc} with strict area and power constraints, such as near-sensor \acp{mcu}. \carus, on the other hand, fits well with those \acp{soc} where a larger area budget can be allocated for increased performance and extended functionalities. The different design goals are mainly reflected in the control strategy adopted in each device. \caesar relies on the host system to stream the necessary \ac{simd} instructions from a reduced \isa. In contrast, \carus integrates a \riscv-based controller supporting a custom \acs{cim}-oriented \isa, paired with a novel and scalable \vpu for maximum flexibility and tunable parallelism at the data level.
Since both architectures are designed for streamlined software deployment, their \isa and microarchitecture were tailored to support standard data types (8-, 16-, and 32-bit integers). Given the limited area budget, support for application-specific lower-precision data types was considered but not implemented, as many edge applications, such as health monitoring and audio processing, still rely on standard data types.

The following sections describe the proposed architectures, motivating their \acp{isa} and microarchitecture.

\subsection{\caesar}
\label{sec:arch-caesar}

\noindent Minimizing the area overhead and power consumption of the \nmc controller and arithmetic circuitry is the top-priority design criteria for the definition of \caesar's \isa and microarchitecture, discussed in \cref{sec:caesar-isa,sec:caesar-micro}.

\subsubsection{\acl*{isa}}
\label{sec:caesar-isa}

\noindent Designed for low-power edge computing applications, \caesar \isa comprises operations that are common in \ac{ml} algorithms such as \acp{cnn} and support vector machines: arithmetic and logic operations that include multiplication, \ac{mac}, dot product, minimum and maximum selection, and shift operations to support fixed-point arithmetic.
A summary of the supported instructions is shown in \cref{tab:caesar-isa}.

\begin{table}[htb]
    \caption{\caesar instruction listing and accumulator operations. All the instructions are element-wise unless otherwise specified.}\label{tab:caesar-isa}
    \scriptsize
    \begin{adjustbox}{width=1\linewidth}
        \begin{tabular}{llll|l|l}
            \toprule
            \multicolumn{4}{c|}{\textbf{Instruction}} &\multicolumn{1}{c|}{\textbf{Acc.}} &\multicolumn{1}{c}{\textbf{Description}}\\
            \midrule
            \multicolumn{6}{c}{\textbf{Arithmetic-Logic Instructions}}\\
            \midrule[0.5pt]
            \texttt{\{AND,OR,XOR\}} &\texttt{dest}  &\texttt{src1}  &\texttt{src2} &-             &Bitwise logic \\
            \texttt{\{ADD,SUB,MUL\}}          &\texttt{dest}  &\texttt{src1}  &\texttt{src2} &-             &Addition, subtraction, mult. \\
            \texttt{MAC\_INIT}    &               &\texttt{src1}  &\texttt{src2} &clear         &Multiply-add initialization \\
            \texttt{MAC}          &               &\texttt{src1}  &\texttt{src2} &update        &Multiply-add \\
            \texttt{MAC\_STORE}   &\texttt{dest}  &\texttt{src1}  &\texttt{src2} &update        &Multiply-add writeback \\
            \texttt{DOT\_INIT}    &               &\texttt{src1}  &\texttt{src2} &clear         &Word-wise dot-product init. \\
            \texttt{DOT}          &               &\texttt{src1}  &\texttt{src2} &update        &Word-wise dot-product \\
            \texttt{DOT\_STORE}   &\texttt{dest}  &\texttt{src1}  &\texttt{src2} &update        &Word-wise dot-product wb. \\
            \texttt{\{SLL,SLR\}}     &\texttt{dest}  &\texttt{src1}  &\texttt{src2} &-             &Logic shift left and right \\
            \texttt{\{MIN,MAX\}}          &\texttt{dest}  &\texttt{src1}  &\texttt{src2} &-             &Min./max. selection \\
            \midrule[0.2pt]
            \multicolumn{6}{c}{\textbf{Configuration Instructions}}\\
            \midrule[0.5pt]
            \texttt{CSRW}         &\multicolumn{3}{l|}{\texttt{bitwidth}}        &-             &Set operand bitwidth in \acs{csr} \\
            \bottomrule
        \end{tabular}
    \end{adjustbox}
\end{table}

When in \emph{computing} mode, \caesar interprets the write transactions on the bus as instructions. The opcode is encoded in the six most significant bits of the data bus, followed by the word address of the two source operands, relative to \caesar base address and 13-bit wide on 32-bit systems, for a total of \SI{32}{\kibi\byte} of addressable space. The address bus encodes the target address (destination address) as in normal accesses. 
For example, suppose that \caesar is mapped at base address \texttt{BASE}, the command to sum together the words at offset \texttt{SRC1} and \texttt{SRC2} and store the result at offset \texttt{DEST} can be assembled and issued online with the following C code:
\centerline{\mintinline[bgcolor=lightgray,fontsize=\footnotesize]{c}{*(BASE + DEST << 2) = ADD << 26 | SRC2 << 13 | SRC1;}}
\noindent Alternatively, an in-house domain-specific compiler can be used to assemble predefined sequences of \caesar instructions that implement specific kernels. These are compiled and embedded into the host system and sent to \caesar by the host \cpu or \ac{dma} controller during execution. To efficiently deal with different data types, all \caesar instructions are packed-\acs{simd}. The data type is statically configured in a \ac{csr} updated by a dedicated instruction to avoid repeated instruction encodings.

\subsubsection{Microarchitecture}
\label{sec:caesar-micro}

\begin{figure}[htb]
    \centering
    \includesvg[width=0.8\linewidth]{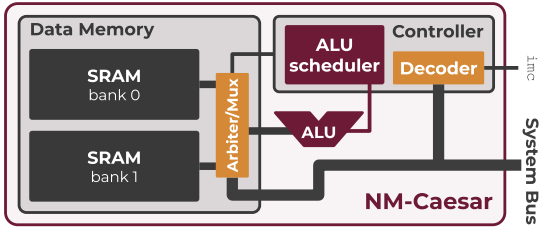}
    \caption{Top-level block diagram of \caesar.}
    \label{fig:caesar_architecture}
\end{figure}

\noindent As shown in \cref{fig:caesar_architecture}, \caesar is composed of two single-port \sram macros, an integer \alu, and a controller that interfaces with the system bus, decodes incoming instructions, and schedules internal memory accesses and arithmetic operations.
Internally, \caesar can access both memory banks simultaneously to increase the bandwidth towards the \alu. Splitting the data memory into two smaller banks has a minor overhead in area and power due to the partially replicated periphery, yet it is more area-efficient than using a single dual-port memory for the same throughput.

\begin{figure}[hbt]
    \centering
    \includegraphics[width=\linewidth]{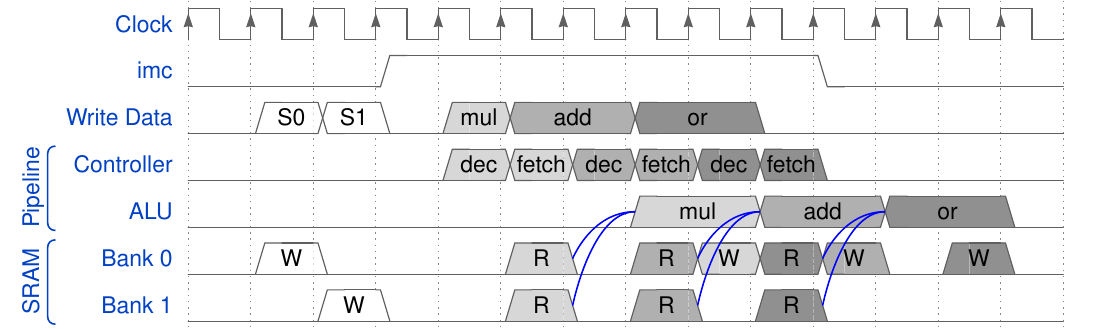}
    \caption{Example timing diagram of \caesar running two normal write operations ($S0$ and $S1$) and three instructions.}
    \label{fig:caesar_timing_diagram}
    \vspace{-4mm}
\end{figure}

The received instructions go through four phases, handled by a 2-stage pipeline as shown in \cref{fig:caesar_timing_diagram}:
\begin{enumerate*}[label=(\arabic*)]
    \item the instruction is decoded (controller \textit{dec} stage), and the source and destination addresses are buffered;
    \item in the next cycle, the source operands are fetched from the two memory banks (controller \textit{fetch} stage);
    \item the operands are received in the following cycle when the necessary \alu operation is scheduled. A new external request can be decoded at this point;
    \item once the \alu provides the result, it is written back to the destination memory location.
\end{enumerate*} 
To minimize the logic area and meet the same timing constraints of the reference \SI{32}{\kibi\byte} \sram, a multi-cycle, 32-bit, \ac{simd} integer \alu was selected. Choosing a single-cycle architecture would not improve performance as read and write contentions would occur at the memory ports.
The \alu design is based on that of the CV32E40P core\cite{riscy} \alu. \caesar inherits its partitioned multi-precision adder supporting 8-, 16-, and 32-bit operations, here implemented with relaxed timing constraints to meet a 2-cycle propagation delay.
The adder is used by the addition, subtraction, and minimum or maximum selection instructions.
Unlike CV32E40P, which employs several parallel multipliers to perform single-cycle \ac{mac} and dot products, \caesar employs four 17-bit multipliers and four adders producing one 8-bit, two 16-bit, and four 8-bit results every two cycles.
Although no data placement constraints exist in \caesar, the throughput is reduced to one operation every three cycles when both source operands come from the same memory bank, where they are accessed sequentially.

\subsection{\carus}
\label{sec:carus-arch}

\noindent \carus architecture is designed to leverage the increased data-access efficiency brought by the \nmc paradigm while also:
\begin{enumerate*}[label=(\arabic*)]
    \item offering the advantages in terms of flexibility, scalability, and instruction fetch overhead that are typically offered by programmable vector machines and\label{item:carus-vector}
    \item relying on a scalable microarchitecture with configurable data-level parallelism\label{item:carus-micro}.
\end{enumerate*}
To achieve \labelcref{item:carus-vector} and \labelcref{item:carus-micro}, \carus integrates an innovative controller based on a small \riscv \cpu, placed near the internal \sram banks and called \ecpu to distinguish it from the external host system's processor. A small memory, the \emem, is programmed by the host system with the \ac{cim} kernel to execute and provides minimal support for stack and heap. The \ecpu offloads vector instructions from the custom \xisa \isa extension presented in \cref{sec:carus-isa} to the novel, area-efficient, scalable \vpu described in \cref{sec:carus-micro}. Unlike \caesar, \carus controller is Turing-complete and fully programmable using any \riscv-compatible software build toolchain. To our knowledge, \carus is the first proposal of a software-programmable, autonomous embedded memory device that supports vector operations through a newly proposed \acs{nmc}-specific \riscv \ac{isa} extension.

Despite integrating a \cpu, \carus remains a memory from a system perspective, following the same system integration model of \caesar. As argued in \cref{sec:related}, the absence of integrated load-store units and master ports in its external interface is what differentiates \carus from a traditional memory-mapped accelerator.
On the other hand, the main difference between \carus and an isolated vector coprocessor with a private memory is the emphasis on the memory capacity over processing resources. The \vrf, which contributes to at least half of \carus die area, is memory-mapped from a system perspective, and it is the only data source for the internal \vpu. In contrast, traditional vector units load and store data from external memories, offering greater flexibility in \vrf data layout at the cost of increased and more expensive data movements.

\subsubsection{\acl*{isa}}
\label{sec:carus-isa}

\noindent The \xisa \riscv custom vector extension is specifically designed for small and efficient \nmc devices to provide sufficient support for parallel workloads while limiting the hardware implementation cost (see \cref{sec:carus-impl}). 
The extension is heavily inspired by the standard \rvv\cite{rvv_spec} and mostly shares the same instruction formats and semantics. \rvv's built-in support for stripmining and very large vectors (up to \SI{64}{\kibi\byte} vectors) perfectly aligns with \carus scalable and memory-centric microarchitecture, allowing the same code to run on implementations with diverse \vrf sizes.

The \xisa custom instructions, listed in \cref{tab:carus-isa}, implement integer single-width arithmetic, logic, and shift operations as well as vector permutation instructions comparable to the standard \rvv counterparts. Besides, \xisa adds two instructions, \mintinline{gas}|xvnmc.emvv| and \mintinline{gas}|xvnmc.emvx|, dedicated to exchanging data between an arbitrary element of a vector register and an arbitrary scalar \ac{gpr} in the \ecpu. This is the only mechanism supported for the \ecpu code to interact with the \vrf content, which is not accessible through normal load and store instructions. Consequently, the \ecpu instructions and private data are stored in the \emem instead.
Vector load and store instructions are not necessary: \carus \vpu works directly in the host system memory space, relying on the system to populate its \vrf.

\begin{table}[htb]
    \centering
    \caption{\xisa instruction listing. Instruction variants are expanded in \cref{tab:carus-formats}. Vector arithmetic operations are element-wise unless otherwise specified. Instruction with the \texttt{[r]} type can optionally use indirect register addressing.}\label{tab:carus-isa}
    \scriptsize
    \begin{minipage}{\linewidth}
    \begin{threeparttable}
        \begin{tabular}{l|ccc|l}
            \toprule
            \multicolumn{1}{c|}{\textbf{Mnemonic}} &\multicolumn{3}{c|}{\textbf{Variants}} &\multicolumn{1}{c}{\textbf{Description}}\\
            \midrule
            \multicolumn{5}{c}{\textbf{Vector Integer Arithmetic-Logic Instructions}}\\
            \midrule[0.5pt]
            \texttt{xvnmc.vadd[r]}             &vv  &vx  &vi &Addition\\
            \texttt{xvnmc.vsub[r]}             &vv  &vx  &   &Subtraction\\
            \texttt{xvnmc.v\{mul,macc\}[r]}    &vv  &vx  &   &Multiplication/MAC\\
            \texttt{xvnmc.v\{and,or,xor\}[r]}  &vv  &vx  &vi &Bitwise logic\\
            \texttt{xvnmc.v\{min,max\}[u][r]}  &vv  &vx  &   &Comparison\\
            \texttt{xvnmc.v\{sll,srl,sra\}[r]} &vv  &vx  &vi &Logic/arith. shift\\
            \midrule[0.2pt]
            \multicolumn{5}{c}{\textbf{Vector Permutation Instructions}}\\
            \midrule[0.5pt]
            \texttt{xvnmc.vmv[r]}                &vv  &vx  &vi &Copy/splat into vector\\
            \texttt{xvnmc.vslide\{up,down\}[r]}  &vx  &vi  &   &Slide elements\\
            \texttt{xvnmc.vslide1\{up,down\}[r]} &vx  &    &   &Slide and push GPR\\
            \midrule[0.2pt]
            \multicolumn{5}{c}{\textbf{Scalar-Vector Movement Instructions}}\\
            \midrule[0.5pt]
            \texttt{xvnmc.emvv}             &ex  &    &   &Move GPR to v[$i$]\\
            \texttt{xvnmc.emvx}             &xe  &    &   &Move v[$i$] to GPR\\
            \midrule[0.2pt]
            \multicolumn{5}{c}{\textbf{Configuration Instructions}}\\
            \midrule[0.5pt]
            \texttt{xvnmc.vset[i]vl[i]}     &\multicolumn{3}{c|}{N/A\tnote{a}} &Set VL and SEW\\
            \bottomrule
        \end{tabular}
        \footnotesize
        \begin{tablenotes}[]
            \item[a] Reserved formats, same as \rvv.
        \end{tablenotes}
    \end{threeparttable}
    \end{minipage}
\end{table}

\begin{table}[htb]
    \centering
    \caption{\carus instruction formats. The \xisa extension is implemented inside the \riscv \emph{Custom-2} 25-bit encoding space with the \texttt{0x5b} major opcode. The standard OPIVV, OPIVX, and OPIVI formats are used for the \texttt{vv}, \texttt{vx}, and \texttt{vi} instruction variants respectively, while \texttt{ex} and \texttt{xe} use the OPMVX format. Indirect register addressing is available for all \texttt{vv}, \texttt{vx}, and \texttt{vi} instructions.}\label{tab:carus-formats}
    \scriptsize
    \begin{tabular}{l|l|ll|l}
        \toprule
        \multicolumn{1}{c|}{\textbf{Var.}} &\multicolumn{1}{c|}{\textbf{Dest.}} &\multicolumn{2}{c|}{\textbf{Data source(s)}} &\multicolumn{1}{c}{\textbf{Example}}\\
        \midrule
        vv   &\texttt{vd[:]}     &\texttt{vs2[:]}    &\texttt{vs1}     &\mintinline{gas}{xvnmc.vadd.vv v2, v1, v0}\\
        vx   &\texttt{vd[:]}     &\texttt{vs2[:]}    &\texttt{rs1}     &\mintinline{gas}{xvnmc.vadd.vx v2, v1, x5}\\
        vi   &\texttt{vd[:]}     &\texttt{vs2[:]}    &\texttt{imm}     &\mintinline{gas}{xvnmc.vadd.vi v2, v1, 42}\\
        ex   &\texttt{vd[rs2]}   &\texttt{-}       &\texttt{rs1}     &\mintinline{gas}{xvnmc.emvv    v0, x4, x5}\\
        xe   &\texttt{rd}        &\texttt{vs2[rs1]}  &\texttt{-}     &\mintinline{gas}{xvnmc.emvx    x5, v0, x4}\\
        \bottomrule
    \end{tabular}
\end{table}

One significant issue when combining a vector \isa without vector loads and stores with the proposed \ac{cim} model is that the layout of the data in the \vrf is not chosen by the vector kernel. As a consequence, $N$ vector instructions with different source and destination registers would be needed to iterate over $N$ chunks, or rows, of the input data mapped to $N$ vector registers. With standard vector instructions, this would translate into a full unrolling of the iteration process. Nested loops in the kernel exacerbate the problem, making the code size grow exponentially. This is not an issue in a traditional vector machine, which can conveniently load data from any memory location into the same vector registers in every loop iteration.
In contrast, the data layout in \carus \vrf is dictated by the host system and thus potentially unknown at compile time if the memory is allocated dynamically. It would not be possible to use the same kernel for different register subsets if the source and destination registers were hardcoded as immediate values in the kernel's vector instructions as in the standard \rvv. 
Moreover, the \emem offers very limited space for the kernel code, whose size is, therefore, a major concern.
To address both of these issues, the \xisa extension provides instruction variants supporting \emph{indirect register addressing}. These variants encode the index of the source and destination vector registers in the three least-significant bytes of a scalar \ac{gpr} (\texttt{rs2}). This solution supports up to 256 logical vectors, accommodating applications with diverse data shapes, similar to \cite{gauchi2020reconfigurable}. As a result, the same vector instruction can be reused in every loop iteration with different operands by simply updating the \ac{gpr} containing their indexes. Consequently, the code size is drastically reduced, and the same kernel can be used for input data with different base addresses and iteration counts (i.e., input data size). This mechanism is equivalent to updating the base address for a load or store instruction in a conventional software loop and, therefore, could easily be integrated into a custom compiler. Moreover, because a single \ac{gpr} encodes all the instruction operands, updating their indexes only requires a single \mintinline{gas}|add| instruction, further reducing the code size. The execution time of index-updating instructions is normally hidden by the latency of the last issued vector instruction, due to \carus ability to execute scalar and vector instructions in parallel. The energy cost of index management instructions is negligible compared to vector operations, as shown by the low contribution of the \ecpu to the total energy in \cref{fig:power-breakdown}.
Finally, the absence of vector load and stores makes the \xisa \isa completely independent of the data bitwidth, further increasing reusability compared to \rvv.

In conclusion, the newly proposed \xisa \isa offers a software-friendly, architecture-agnostic interface to leverage the proposed \ac{cim} model, bridging the host system's \emph{memory} view of the \ac{cim} device with a convenient internal \emph{register} view. Through vector processing, data-level parallelism and energy efficiency are increased, while a smaller code size and increased reusability compared to conventional vector \acp{isa} are achieved using indirect vector register addressing.

From an application standpoint, the interaction between the host \cpu and the \carus instance is implemented through a driver that allows developers to program the \emem inside the controller with a \xisa program selected from a \emph{library} of precompiled kernels. Although support for the \xisa extension is currently available only at the assembler level, its compatibility with the standard \riscv \isa and its similarity to the standard \rvv facilitate potential integration at the compiler level for automatic kernel generation starting from high-level languages.

\subsubsection{Microarchitecture}
\label{sec:carus-micro}

\begin{figure}[htb]
    \centering
    \includesvg[width=\linewidth]{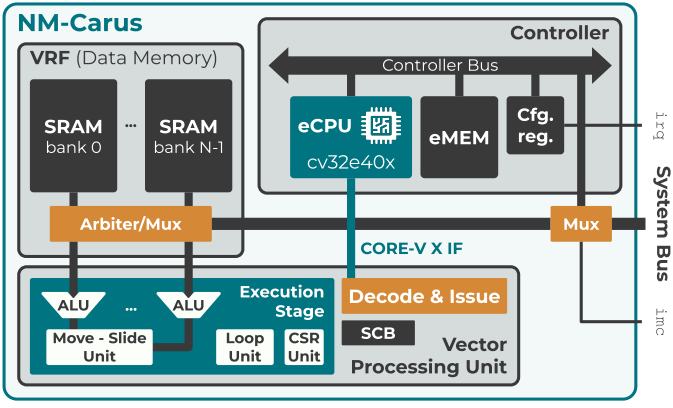}
    \caption{Top-level block diagram of \carus.}
    \label{fig:carus-top}
    \vspace{-4mm}
\end{figure}

\noindent A block diagram of \carus microarchitecture is presented in \cref{fig:carus-top}.

\begin{figure}[h]
    \centering
    \includegraphics[width=\linewidth]{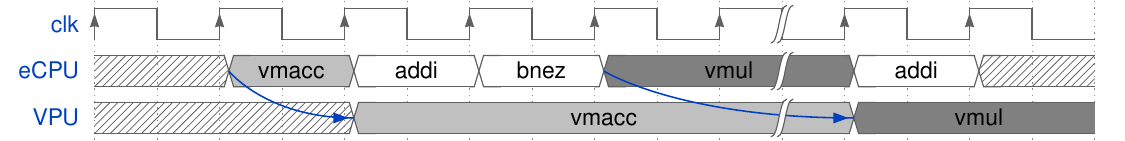}
    \caption{Scalar and vector instruction execution in \carus.}
    \label{fig:carus-timing}
\end{figure}

\carus controller is essentially a minimal \soc featuring the OpenHW Group 4-stage, in-order CV32E40X~\cite{cv32e40x} \riscv core\footnote{Smaller \riscv \acp{cpu} exist and would better suit \carus controller. CV32E40X was chosen because of its CORE-V X interface that allows straightforward implementation of custom coprocessors.} (\acs{ecpu}), a tiny code memory (\acs{emem}), and a configuration register that implements the synchronization interface with the host system. To minimize the core area, the \ecpu implements the RV32EC \isa, offering 16 \acp{gpr} and no hardware multiplication and division. Instructions from the \xisa extension are offloaded by the \ecpu to the \vpu through the CORE-V X interface\cite{xif}. All controller components are interconnected through a single-channel bus that is exposed to the host as a memory-mapped slave peripheral through the multiplexed memory-like interface that is also used to access \carus data memory. In \emph{configuration} mode, the host can program the \emem with the kernel code or trigger and check the kernel execution by accessing the configuration register. \carus can be set back to normal \emph{memory} mode during the kernel execution so that normal memory operations are possible (e.g., to implement double buffering).
Once the kernel terminates, a dedicated status bit is set to signal the end of the computation to the host system. As an alternative to software polling, this bit is routed to a dedicated optional pin in the \carus top-level interface, which may be used as an interrupt source to allow the host system to enter a low-power state during computation.
As explained in \cref{sec:carus-isa}, \mintinline{gas}|xvnmc.emvx| instructions are the only ones that cause data hazards between vector and scalar instructions. Therefore, in most cases, scalar and vector instructions can be executed in parallel as shown in \cref{fig:carus-timing}, reducing the kernel execution time.

\carus \vpu is implemented as a single-issue vector machine with a configurable level of hardware unrolling. The pipeline is composed of three main parts:
\begin{enumerate*}[label=(\arabic*)]
    \item a \emph{decode stage} interfacing with the \ecpu and issuing decoded controls and scalar data to the execution units,
    \item an \emph{execution stage} with three dedicated units handling arithmetic, permutation, and \ac{csr} instructions, and
    \item a shared \emph{commit unit} keeping track of the execution status of the two in-flight instructions through a minimal scoreboard and handling synchronization with the \ecpu.
\end{enumerate*}
The entire \vpu is clock-gated when no vector instructions are in flight. The rationale behind a single-issue design is to keep the control and arbitration logic and internal buffers as small as possible. Consequently, most of the available area and power budget is allocated to the arithmetic circuitry, which ultimately determines the throughput of the \vpu. A wider or deeper execution pipeline typical of performance-oriented vector machines \cite{cavalcante2020ara} would have guaranteed a greater utilization of the functional units and \vrf data interfaces. However, because \carus is intended to be configured with a large \ac{vrf}, the idle time of the execution units is negligible compared to the total execution time of each instruction, which fails to justify the additional hardware and power cost required to minimize it.
\carus data memory is implemented as a configurable number of single-port, 32-bit \sram banks with arbitrary capacity.
The current \vrf architecture supports 32 logical vector registers like the standard \rvv.
The chosen data interleaving policy, shown in \cref{fig:carus-vrf}, maps words that are contiguous inside the host system address space to adjacent \vrf banks. All vector registers are naturally aligned with the physical banks, so elements with the same index from different vectors are physically mapped to the same \vrf bank. This enables straightforward vector unrolling, with all the operands of an element-wise instruction being mapped to the same bank. Therefore, each \alu in the arithmetic unit is connected to a single \sram bank, forming an independent computing \textit{lane}. Consequently, \carus \vpu can be scaled arbitrarily: a higher number of lanes increases the unrolling level, thus improving throughput at the cost of increased area and power consumption.
Because all operands come from the same bank, they must be accessed sequentially. This does not represent a bottleneck since \carus relies on small serial arithmetic circuits, offering compelling advantages from an implementation point of view compared to parallel implementations. Providing the necessary throughput to parallel arithmetic units would also require larger and less efficient multi-port memory banks or more complex \vrf layouts with additional routing and hazard-detection hardware.
In addition, the supported vector instructions range from a minimum of one vector source operand (e.g., \mintinline{gas}|xvnmc.vadd.vx|) per result to a maximum of three (e.g., \mintinline{gas}|xvnmc.vmacc.vv|). Optimizing the memory access patterns for such a wide range of operations would exacerbate the issue.

\begin{figure}[htb]
    \centering
    \includesvg[width=0.6\linewidth]{carus-vrf.svg}
    \caption{\carus \acl{vrf} organization.}
    \label{fig:carus-vrf}
    \vspace{-2mm}
\end{figure}

At the core of \carus \vpu is an execution engine with three dedicated execution units:
\begin{enumerate*}[label=(2.\alph*)]
    \item an \emph{arithmetic unit} handling integer arithmetic and logic instructions,\label{item:carus-arith}
    \item a \emph{move-slide unit} performing permutation and scalar-vector movement operations, and\label{item:carus-move}
    \item a \emph{\acs{csr} unit} managing the access to the \texttt{vtype} \ac{csr} encoding the current vector length and element bitwidth.
\end{enumerate*}
A shared loop unit generates the \vrf addresses to iterate among the elements of the source and destination vectors. The datapaths of \labelcref{item:carus-arith} and \labelcref{item:carus-move} are interdependent since the \acp{alu} inside the arithmetic unit and their internal registers are exploited to buffer intermediate data from both arithmetic and permutation operations. Similarly, resource sharing is heavily exploited inside the \acp{alu} to implement packed-\acs{simd} operations with minimal area. To this purpose, each \alu is equipped with a partitioned multi-precision 16-bit adder (similar to \caesar's one), a 16-bit multiplier, elementary logic operators, and a serial 8-bit logic and arithmetic barrel shifter.
32-bit addition is performed as two successive 16-bit additions, 32-bit multiplication is computed with three successive 16-bit multiplications accumulated by the adder, and 8-bit multiplications are sign-extended to 16 bits and truncated. As a result, the adder can process 32-bit data words every two cycles regardless of the element bitwidth, while the multiplier produces four 8-bit, two 16-bit, and one 32-bit results in four, two, and three cycles respectively.

Every \ac{alu} performs the same operation on different elements of the same vector. Therefore, all lanes share the same \alu control unit and memory access scheduler. The compute and memory access patterns ensure that the throughput of the arithmetic unit is never lower than the slower unit between the \alu and the \vrf for any combination of \alu operation and operand bitwidth. Most notably, an iteration of the \mintinline{gas}|xvnmc.vmacc.vx| instruction at the core of many linear algebra operations is executed in each lane with a throughput of \SI{1}{MAC \per cycle}, \SI{0.67}{MAC \per cycle}, and \SI{0.33}{MAC \per cycle} for 8-bit, 16-bit, and 32-bit operands, respectively.

%% file: implementation.tex
\section{Physical Implementation}
\label{sec:impl}

\begin{figure}[hbt]
    \centering
    \includesvg[width=\linewidth]{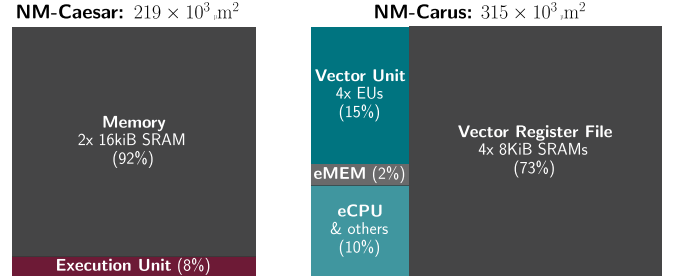}
    \caption{Post-synthesis area breakdown of \SI{32}{\kibi\byte} instances of \caesar and \carus in accurate scale.}
    \label{fig:nmc_breakdown}
    \vspace{-4mm}
\end{figure}

\noindent Both \caesar and \carus have been implemented in a \SI{32}{\kibi\byte} configuration on a low-power \SI{65}{\nano\meter} CMOS technology library.
The automated digital flow relies on Synopsys Design Compiler\textsuperscript{\textregistered} 2020.09 for logic synthesis and Cadence Innovus\textsuperscript{\textregistered} 20.1 for place and route. To meet the design goal of implementing two drop-in replacements for standard \sram memories, the cycle time and input/output delay constraints under worst-case conditions were set to match those of a single-port reference \SI{32}{\kibi\byte} memory macro generated by the same \sram compiler used for internal \sram banks.
The same metal stack of standard \sram macros was used, which favours little to no effort when integrating the \nmc macros in a host system.
The post-layout area and timing characteristics of the \nmc macros are summarized in \cref{tab:implementation_table}, while the post-synthesis area breakdown is shown in \cref{fig:nmc_breakdown}. The following sections discuss the implementation details of each \nmc \ac{ic}.

\begin{table}[htb]
    \centering
    \caption{Post-layout area and timing characteristics of a \SI{32}{\kibi\byte} traditional \sram, \caesar, and \carus when implemented on a low-power \SI{65}{\nano\meter} CMOS technology node.}
    \label{tab:implementation_table}
    \scriptsize
    \begin{tabular}{l|rrr}
        \toprule
        \begin{tabular}[c]{@{}c@{}}\textbf{Metric}\end{tabular} & \textbf{\sram} & \textbf{\caesar}                                                 & \textbf{\carus}                                                    \\ \midrule
        \begin{tabular}{@{}l@{}}Post-layout\\area [\SI{}{\micro\meter\squared}]\end{tabular}   & \num{200e3} & \begin{tabular}[c]{@{}r@{}}\num{256e3}\\ (\SI{+28}{\percent})\end{tabular} & \begin{tabular}[c]{@{}r@{}}\num{419e3}\\ (\SI{+110}{\percent})\end{tabular}           \\ \midrule
        \begin{tabular}{@{}l@{}}Post-layout max\\clock freq. [\SI{}{\mega\hertz}]\end{tabular}          & 330  & 330                                                       & 330                                                                     \\ \midrule
        \begin{tabular}{@{}l@{}}Max input\\delay [\SI{}{\nano\second}]\end{tabular} & 0.69  & \begin{tabular}[c]{@{}r@{}}0.70\\ (\SI{+2}{\percent})\end{tabular}      & \begin{tabular}[c]{@{}r@{}}0.70\\ (\SI{+2}{\percent})\end{tabular}                   \\ \midrule
        \begin{tabular}{@{}l@{}}Max output\\delay [\SI{}{\nano\second}]\end{tabular} & 2.28  & \begin{tabular}[c]{@{}r@{}}2.28\\ (\SI{+0}{\percent})\end{tabular} & \begin{tabular}[c]{@{}r@{}}2.48\\ (\SI{+9}{\percent})\end{tabular}                  \\ \bottomrule
    \end{tabular}
\end{table}

\subsection{\caesar}
\noindent In a \SI{2}{\times} \SI{16}{\kibi\byte} configuration for its internal memories, \caesar meets the target \SI{330}{\mega\hertz} clock frequency of the reference \SI{32}{\kibi\byte} \sram bank. The output delay is unchanged, while the worst-case input delay is \SI{2}{\percent} higher. The post-layout area is \qtyproduct{673 x 380}{\micro\meter\squared}, corresponding to an overhead of \SI{28}{\percent} over the reference \SI{32}{\kibi\byte} \sram. \Cref{fig:caesar_floorplan} shows the floorplan of \caesar, highlighting its main internal components. The two memory banks are placed on the top of the design to minimize the distance between the logic and the interface pins in the lower part of the \ac{ic}. This ensures the same pin orientation of the reference \sram bank, easing the system-level routing phase while relaxing the timing constraints.

\begin{figure}[htb]
    \centering
    \centering
    \includesvg[width=0.7\linewidth]{caesar-layout.svg}
    \caption{\SI{32}{\kibi\byte} \caesar layout.}
    \label{fig:caesar_floorplan}
    \vspace{-5mm}
\end{figure}

\subsection{\carus}
\label{sec:carus-impl}

\noindent \carus has been implemented in a four-lane configuration, with the controller and \vpu placed in the middle of four \SI{8}{\kibi\byte} \ac{sram} macros constituting the internal \vrf as shown in \cref{fig:carus_floorplan}. The \emem is implemented as a \SI{512}{\byte} register file macro. The target \SI{330}{\mega\hertz} clock frequency is met while the input and output delays are increased by \SI{2}{\percent} and \SI{9}{\percent} respectively due to the additional multiplexing logic that connects either the correct \vrf bank or the controller bus to the shared external bus interface.
The post-layout area is \qtyproduct{705.0 x 594.4}{\micro\meter\squared}, approximately corresponding to twice the area of the reference \SI{32}{\kibi\byte} \ac{sram}, therefore meeting the target \SI{50}{\percent} memory to logic ratio.
Although most of the increased area overhead compared to \caesar comes from \carus additional logic, it is partially determined by the sublinear scaling of the footprint of an \sram with its reduction in size. As visible in \cref{fig:nmc_breakdown}, \carus 4-bank data memory is larger than \caesar's 2-bank one, despite their identical capacity.

Placing the \vrf banks at the corner of the \carus floorplan partially counteracts the routing congestion, which is the main reason behind the low \SI{60}{\percent} logic density.
With this solution, the \vpu vector permutation unit, which interconnects all the \vrf banks, is placed in the center of the \ac{ic}, minimizing the wire length.
Extra space is left between adjacent memories to route the \carus input and output pins to the middle logic part, limiting timing penalties while keeping the pins on the bottom side as in the reference macro. Implementations based on more advanced technology nodes would benefit from the higher scaling of logic compared to the \sram arrays, reducing the area overhead or allowing for a higher lane count for a given area budget.

\begin{figure}[htb]
    \centering
    \centering
    \includesvg[width=0.7\linewidth]{carus-layout.svg}
    \caption{\SI{32}{\kibi\byte} \carus layout (4 lanes).}
    \label{fig:carus_floorplan}
    \vspace{-4mm}
\end{figure}


%% file: performance.tex
\section{Performance and Energy Assessment}
\label{sec:perf}


\subsection{Experimental Setup}

\subsubsection{System Integration}
\noindent The two \SI{32}{\kibi\byte} \nmc macros presented in \cref{sec:impl} were integrated inside X-HEEP\cite{machetti2024xheep}, an open-source, configurable and low-power \riscv \mcu. Two of the eight \SI{32}{\kibi\byte} \sram banks connected to the X-HEEP bus were replaced by \caesar and \carus, and the necessary configuration registers were added to the peripheral subsystem of X-HEEP. \carus interrupt pin was routed to the system \cpu, the OpenHW Group CV32E40P in-order 4-stages \riscv processor implementing the RV32IMC extensions. The necessary input data is directly embedded in the firmware and loaded into the \nmc macros at startup to emulate a real-world online data acquisition and processing scenario.
Using worst-case timing constraints, the system was synthesized, placed, and routed on a low-power \SI{65}{\nano\meter} technology library. All energy data presented in this section was collected from power analysis performed in Synopsys Primepower\textsuperscript{\textregistered} 2019.12 under typical operating conditions using \acp{vcd} from system-level post-layout simulations when running selected benchmark applications.

\begin{figure}[hbt]
    \centering
    \includesvg[width=\linewidth]{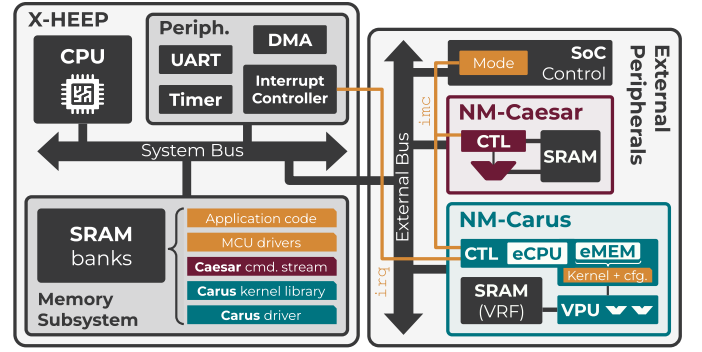}
    \caption{Block diagram and code placement of the \heeperator system used to evaluate \caesar and \carus performance.}
    \label{fig:heeperator}
\end{figure}

\subsubsection{Benchmarks}
\noindent To assess the system-level execution time and energy efficiency gains unlocked by the proposed \nmc architectures, a set of representative computing kernels was selected and implemented in a \cpu-only version used as a baseline and two \nmc-enhanced variants for \caesar and \carus. The chosen kernels, listed in \cref{tab:carus-caesar-kernels,fig:energy-bench}, constitute the basic building block for more complex applications that modern edge devices are expected to run, ranging from simple bitwise or element-wise arithmetic-logic operations to linear-algebra (matrix multiplication, 2D convolution) and machine-learning-specific kernels (\acs{relu}, max pooling).
\caesar instruction sequences are generated using a dedicated in-house domain-specific compiler and streamed from memory to the \nmc macro by the system \ac{dma} controller. 
\carus kernels are written in C or \riscv assembly, compiled using an extended version of the GNU \riscv GCC compiler with assembler support for the \xisa vector extension, and copied from memory to the \emem at run time. Finally, the \cpu-only kernels are compiled for the low-power system \cpu using the \riscv RV32IMC instruction set.
To ensure consistency when comparing the cycle count, all the code is compiled with the \texttt{-O3} optimization flag, using the same GNU \riscv GCC compiler, version 11.1.0.

\subsection{Experimental Results}
\label{sec:results}
\subsubsection{Recurrent kernels}
\noindent \Cref{tab:carus-caesar-kernels} reports the relative execution time and energy reduction of the \nmc-enhanced benchmark kernels compared to their reference \cpu-only implementation.
The speedup in arithmetic and linear algebra kernels ranges from \SI{3.3}{\times} to \SI{28.0}{\times} for \caesar and from \SI{6.6}{\times} to \SI{53.9}{\times} for \carus, depending on the data width.
Simpler operations, such as bitwise XOR and element-wise addition, show no significant improvement when switching to lower-precision 16- and 8-bit data types. This is because the compiler's auto-vectorization optimizes the \cpu code by packing multiple data items into a single \ac{gpr}, resulting in a linear performance increase as data width decreases—similar to what occurs in the \nmc devices.
The execution time is further reduced up to \SI{99.6} times in the case of \ac{relu} in \carus, thanks to the inclusion of minimum and maximum selection instructions in its \xisa \isa, which brings a significant speedup compared to the \cpu implementation relying on inefficient data-dependent branches instead. On the other hand, the \nmc architectures only achieve marginal improvement in the execution time in the max pooling kernel. The lack of sub-word reduction operations in \caesar and vector reduction operations in \carus requires horizontal pooling to be implemented in software in the system \cpu and on \carus \ecpu. In general, the energy efficiency of the proposed \nmc \acp{ic} follows the same trend as the throughput, with significant gains over the \cpu baseline, as better visualized in \cref{fig:energy-bench}. It must be noted that the performance gain when moving to sub-word data sizes is uniquely due to the packed-\ac{simd} architecture of the arithmetic units inside the \nmc macros. The same approach has also been proven to be effective in conventional computing systems, typically in the form of dedicated \isa extensions such as \cite{riscy}.

\begin{table*}[htb]
    \caption{System-level throughput and energy improvement of the \heeperator system when executing the same kernels on \caesar and \carus, relative to their \acs{cpu}-only implementation (baseline). Improvements are computed as \cpu data divided by \caesar or \carus data (higher is better). Data source: post-layout simulation (\SI{65}{\nano\meter}, $f_{CLK} = \SI{250}{\mega\hertz}$).}
    \label{tab:carus-caesar-kernels}
    \begin{subtable}{1\linewidth}
        \centering
        \begin{adjustbox}{width=1\linewidth}
        \begin{threeparttable}
        \begin{tabular}{ll|r|r|r|r|r|r|r|r|r|r|}
        \toprule
        \multicolumn{2}{c|}{\textbf{Device and bitwidth}} & \multicolumn{2}{c|}{\textbf{Bitwise XOR}\tnote{a}}                     & \multicolumn{2}{c|}{\textbf{Element-wise addition}\tnote{a}}                             & \multicolumn{2}{c|}{\textbf{Element-wise multiplication}\tnote{a}}                            & \multicolumn{2}{c|}{\textbf{Matrix multiplication}\tnote{b}}                       & \multicolumn{2}{c|}{\textbf{\acs{gemm}}\tnote{c}}                            \\ \midrule
        \multicolumn{2}{c|}{\textbf{Baseline}}          & \multicolumn{1}{c|}{\textbf{\textbf{Cycles/output}}}    & \textbf{\textbf{Energy/output}}         & \multicolumn{1}{c|}{\textbf{Cycles/output}}    & \textbf{Energy/output}         & \multicolumn{1}{c|}{\textbf{Cycles/output}}    & \textbf{Energy/output}         & \multicolumn{1}{c|}{\textbf{Cycles/output}}    & \textbf{Energy/output}         & \multicolumn{1}{c|}{\textbf{Cycles/output}}    & \textbf{Energy/output}         \\ \midrule
        \multirow{3}{*}{\begin{tabular}{l}\riscv \cpu\\(RV32IMC)\end{tabular}} & 8-bit & \num{2.5}           &     \SI{61}{\pico\joule}     & \num{4.0}           &    \SI{99}{\pico\joule}      & \num{11.0}           &     \SI{267}{\pico\joule}     & \num{112.0}           &      \SI{2.88}{\nano\joule}   & \num{73.1}           &      \SI{1.91}{\nano\joule}   \\ 
        {} & 16-bit & \num{5.0}            &     \SI{124}{\pico\joule}     & \num{11.0}            &    \SI{269}{\pico\joule}      & \num{11.0}            &     \SI{285}{\pico\joule}     & \num{112.0}           &      \SI{3.00}{\nano\joule}   & \num{81.2}           &      \SI{2.26}{\nano\joule}      \\ 
        {} & 32-bit & \num{10.0}            &     \SI{281}{\pico\joule}     & \num{10.0}            &    \SI{278}{\pico\joule}      & \num{10.0}            &     \SI{279}{\pico\joule}     & \num{89.1}           &      \SI{2.54}{\nano\joule}   & \num{66.3}           &      \SI{1.95}{\nano\joule}   \\ \midrule[0.2pt]
        \multicolumn{2}{c|}{\textbf{Improvement}}     & \multicolumn{1}{c|}{\textbf{Throughput}}     & \multicolumn{1}{c|}{\textbf{Energy}}         & \multicolumn{1}{c|}{\textbf{Throughput}}     & \multicolumn{1}{c|}{\textbf{Energy}}         & \multicolumn{1}{c|}{\textbf{Throughput}}     & \multicolumn{1}{c|}{\textbf{Energy}}         & \multicolumn{1}{c|}{\textbf{Throughput}}     & \multicolumn{1}{c|}{\textbf{Energy}}         & \multicolumn{1}{c|}{\textbf{Throughput}}     & \multicolumn{1}{c|}{\textbf{Energy}}         \\ \midrule[0.5pt]
        \multirow{3}{*}{\begin{tabular}{l}\caesar\\ (\SI{32}{\kibi\byte})\end{tabular}} & 8-bit   & \SI{5.0}{\times} & \SI{4.0}{\times} & \SI{8.0}{\times} & \SI{6.4}{\times} & \SI{22.0}{\times} & \SI{17.4}{\times} & \textbf{\SI{28.0}{\times}} & \textbf{\SI{25.0}{\times}} & \textbf{\SI{9.1}{\times}} & \textbf{\SI{8.1}{\times}} \\ 
        {} & 16-bit  & \SI{5.0}{\times}          & \SI{4.1}{\times}             & \SI{11.0}{\times}          & \SI{8.9}{\times}             & \SI{11.0}{\times}          & \SI{9.5}{\times}           & \SI{14.0}{\times}          & \SI{13.4}{\times}          & \SI{6.7}{\times}           & \SI{6.5}{\times}           \\ 
        {} & 32-bit  & \SI{5.0}{\times}           & \SI{4.7}{\times}           & \SI{5.0}{\times}           & \SI{4.7}{\times}           & \SI{5.0}{\times}           & \SI{4.7}{\times}           & \SI{5.6}{\times}           & \SI{5.8}{\times}           & \SI{3.3}{\times}           & \SI{3.4}{\times}           \\ \midrule[0.2pt]
        \multirow{3}{*}{\begin{tabular}{l}\carus\\ (\SI{32}{\kibi\byte},\\4 lanes)\end{tabular}} & 8-bit & \SI{12.7}{\times} & \SI{6.6}{\times} & \SI{20.3}{\times} & \SI{10.6}{\times} & \SI{42.0}{\times}   & \SI{23.7}{\times} & \textbf{\SI{53.9}{\times}}   & \textbf{\SI{35.6}{\times}} & \textbf{\SI{31.6}{\times}} & \textbf{\SI{20.7}{\times}} \\ 
        {} & 16-bit   & \SI{12.7}{\times}          & \SI{6.7}{\times}          & \SI{27.9}{\times}          & \SI{14.5}{\times}          & \SI{27.9}{\times}          & \SI{14.9}{\times}          & \SI{37.1}{\times}          & \SI{21.8}{\times}          & \SI{24.1}{\times}          & \SI{14.4}{\times}          \\ 
        {} & 32-bit  & \SI{12.7}{\times}          & \SI{7.5}{\times}           & \SI{12.7}{\times}          & \SI{7.5}{\times}           & \SI{12.6}{\times}          & \SI{7.1}{\times}           & \SI{11.0}{\times}          & \SI{7.1}{\times}           & \SI{7.3}{\times}           & \SI{4.8}{\times}           \\ \bottomrule
        \end{tabular}
        \footnotesize
        \begin{tablenotes}[]
            \item [a] Input data size: \SI{8}{\kibi\byte} (\caesar), \SI{10}{\kibi\byte} (\cpu and \carus).
            \item [b] $A[8,8] \times B[8,p]$, with $p=\{128,256,512\}$ (\caesar) and $p=\{256,512,1024\}$ (\cpu and \carus) for $\{32,16,8\}$ bits.
            \item [c] $\alpha \left(A[8,8] \times B[8,p]\right) + \beta C[8,p]$, with $p=\{128,256,512\}$ (\caesar) and $p=\{256,512,1024\}$ (\cpu and \carus) for $\{32,16,8\}$ bits.
        \end{tablenotes}
        \end{threeparttable}
        \end{adjustbox}
    \end{subtable}
    
    \bigskip
    
    \begin{subtable}{1\linewidth}
        \centering
        \begin{adjustbox}{width=0.82\linewidth}
        \begin{threeparttable}
        \begin{tabular}{ll|r|r|r|r|r|r|r|r|}
        \toprule
        \multicolumn{2}{c|}{\textbf{Device and bitwidth}} & \multicolumn{2}{c|}{\textbf{2D convolution}\tnote{d}}                        & \multicolumn{2}{c|}{\textbf{ReLU}\tnote{e}}                            & \multicolumn{2}{c|}{\textbf{Leaky ReLU}\tnote{e,f}}                      & \multicolumn{2}{c|}{\textbf{Maxpooling}\tnote{g}}                       \\ \toprule
        \multicolumn{2}{c|}{\textbf{Baseline}}          & \multicolumn{1}{c|}{\textbf{Cycles/output}}    & \textbf{Energy/output}         & \multicolumn{1}{c|}{\textbf{Cycles/output}}    & \textbf{Energy/output}         & \multicolumn{1}{c|}{\textbf{Cycles/output}}    & \textbf{Energy/output}         & \multicolumn{1}{c|}{\textbf{Cycles/output}}   & \textbf{Energy/output}        \\ \midrule
        \multirow{3}{*}{\begin{tabular}{l}\riscv \cpu\\(RV32IMC)\end{tabular}} & 8-bit & \num{135.0}           &   \SI{3.3}{\nano\joule}       & \num{13.0}           &     \SI{344}{\pico\joule}     & \num{12.0}           &     \SI{300}{\pico\joule}     & \num{64.6}           &    \SI{1.44}{\nano\joule}    \\ 
        {} & 16-bit & \num{133.0}           &   \SI{3.4}{\nano\joule}       & \num{12.0}            &      \SI{338}{\pico\joule}    & \num{11.5}            &     \SI{295}{\pico\joule}     & \num{65.6}           &     \SI{1.5}{\nano\joule}    \\ 
        {} & 32-bit & \num{115.1}           &   \SI{3.1}{\nano\joule}       & \num{10.0}            &      \SI{300}{\pico\joule}    & \num{9.5}            &     \SI{258}{\pico\joule}     & \num{50.3}           &     \SI{1.2}{\nano\joule}    \\ \midrule[0.2pt]
        \multicolumn{2}{c|}{\textbf{Improvement}}     & \multicolumn{1}{c|}{\textbf{Throughput}}     & \multicolumn{1}{c|}{\textbf{Energy}}         & \multicolumn{1}{c|}{\textbf{Throughput}}     & \multicolumn{1}{c|}{\textbf{Energy}}         & \multicolumn{1}{c|}{\textbf{Throughput}}     & \multicolumn{1}{c|}{\textbf{Energy}}         & \multicolumn{1}{c|}{\textbf{Throughput}}     & \multicolumn{1}{c|}{\textbf{Energy}} \\ \midrule[0.5pt] 
        \multirow{3}{*}{\begin{tabular}{l}\caesar\\ (\SI{32}{\kibi\byte})\end{tabular}} & 8-bit & \textbf{\SI{16.9}{\times}} & \textbf{\SI{14.2}{\times}} & \textbf{\SI{26.0}{\times}} & \textbf{\SI{22.4}{\times}} & \SI{12.0}{\times} & \SI{10.3}{\times}    & \SI{3.9}{\times}   & \SI{3.8}{\times} \\ 
        {}  & 16-bit & \SI{8.3}{\times}           & \SI{7.6}{\times}           & \SI{12.0}{\times}          & \SI{11.6}{\times}          & \SI{5.7}{\times}           & \SI{5.0}{\times}           & \SI{3.5}{\times}          & \SI{3.5}{\times}          \\ 
        {}  & 32-bit & \SI{6.4}{\times}           & \SI{6.1}{\times}           & \SI{5.0}{\times}           & \SI{5.1}{\times}           & \SI{2.4}{\times}           & \SI{2.2}{\times}           & \SI{6.1}{\times}          & \SI{5.8}{\times}          \\ \midrule[0.2pt]
        \multirow{3}{*}{\begin{tabular}{l}\carus\\ (\SI{32}{\kibi\byte},\\4 lanes)\end{tabular}} & 8-bit & \textbf{\SI{47.5}{\times}} & \textbf{\SI{29.4}{\times}} & \textbf{\SI{99.6}{\times}} & \textbf{\SI{59.3}{\times}} & \SI{26.9}{\times} & \SI{17.3}{\times} & \SI{6.3}{\times} & \SI{6.7}{\times} \\ 
        {}   & 16-bit & \SI{29.3}{\times}          & \SI{17.6}{\times}          & \SI{46.0}{\times}            & \SI{28.9}{\times}          & \SI{12.9}{\times}          & \SI{8.6}{\times}           & \SI{5.7}{\times}          & \SI{5.8}{\times}          \\ 
        {}   & 32-bit & \SI{10.0}{\times}          & \SI{6.3}{\times}           & \SI{19.1}{\times}          & \SI{2.8}{\times}          & \SI{5.3}{\times}           & \SI{3.7}{\times}           & \SI{3.7}{\times}          & \SI{3.5}{\times}          \\ \bottomrule
        \end{tabular}
        \begin{tablenotes}[]
            \item [d] $A[8,n] \circledast F[f,f]$, with $n=\{64,64,128\}$, $f=\{3,4,4\}$ (\caesar) and $n=\{256,512,1024\}$, $f=3$ (\cpu and \carus) for $\{32,16,8\}$ bits.
            \item [e] Input data size: \SI{8}{\kibi\byte} (\caesar), \SI{16}{\kibi\byte} (\cpu and \carus).
            \item [f] Negative slope coefficient implemented as right shift (only powers of 2).
            \item [g] Pooling window: $W\left[2,2\right]$ (stride: 2). Data size: \SI{8}{\kibi\byte} (\caesar), \SI{16}{\kibi\byte} (\cpu and \carus). \caesar version partially implemented on the \cpu.
        \end{tablenotes}
        \end{threeparttable}
        \end{adjustbox}
    \end{subtable}
\end{table*}

\begin{figure*}[htb]
    \centering
    \includesvg[width=\linewidth]{energy-bench.svg}
    \caption{Energy efficiency gain of the \nmc-enhanced \heeperator \mcu compared with its \cpu-only version. Data source: post-layout simulation (\SI{65}{\nano\meter}, $f_{CLK} = \SI{250}{\mega\hertz}$).}
    \label{fig:energy-bench}
    \vspace{-4mm}
\end{figure*}

\begin{figure*}[htb]
    \centering
    \begin{subfigure}{.5\textwidth}
        \centering
        \includesvg[width=0.8\linewidth]{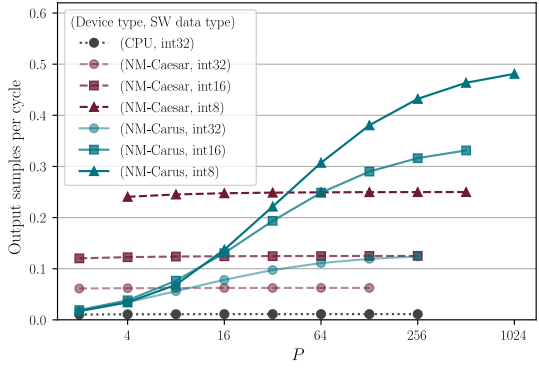}
        \caption{Matrix multiplication ($\left[8,8\right] \times \left[8,P\right]$) throughput scaling.}
        \label{fig:thr-scaling-matmul}
    \end{subfigure}%
    \begin{subfigure}{.5\textwidth}
        \centering
        \includesvg[width=0.8\linewidth]{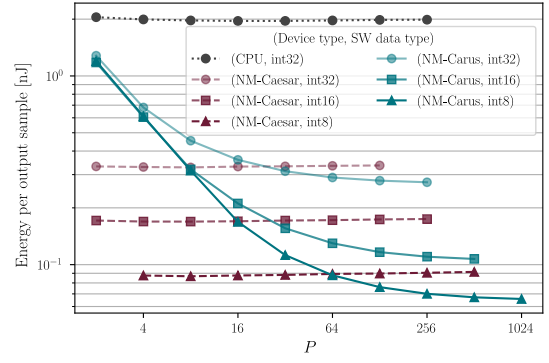}
        \caption{Matrix multiplication ($\left[8,8\right] \times \left[8,P\right]$) energy scaling.}
        \label{fig:energy-scaling-matmul}
    \end{subfigure}
    \caption{Throughput \subref*{fig:thr-scaling-matmul} and energy \subref*{fig:energy-scaling-matmul} scaling of \caesar (red) and \carus (cyan) compared to the \cpu-only \mcu version (grey) on several matrix multiplication kernels with different shape and bitwidth. The \cpu throughput does not vary significantly with data size when using the standard RV32IMC \isa, therefore only 32-bit data is shown. Post-layout simulation data. Driver overhead not considered.}
    \label{fig:scaling-matmul}
    \vspace{-2mm}
\end{figure*}

The data in \cref{tab:carus-caesar-kernels} was collected during kernel runs over large input datasets to show the maximum potential of the proposed architectures. However, to better demonstrate and motivate the architectural differences between \caesar and \carus, their performance was also measured when executing a matrix multiplication kernel on input matrices with varying dimensions. The results of this analysis, reported in \cref{fig:scaling-matmul}, highlight two significantly different scaling trends. \caesar throughput (\cref{fig:thr-scaling-matmul}) and energy efficiency (\cref{fig:energy-scaling-matmul}) improvements over the \cpu baseline remain constant with the input size. Offloading workloads to \caesar has a negligible overhead of five cycles, making it effective even with short computing tasks. In contrast, the overhead of \carus \cpu-based controller hinders its performance with small workloads, because the \ecpu always executes a small sequence of instructions to bootstrap the offloaded kernel. Besides, as pointed out in \cref{sec:carus-micro}, its single-issue, in-order \vpu cannot guarantee optimal utilization of its \acp{alu} between consecutive instructions since a limited number of \alu idle cycles is experienced while waiting for the writeback of the previous vector instruction to complete. Such overhead becomes negligible for larger data sizes, allowing \carus to offer higher throughput and energy efficiency than \caesar, saturating at \SI{0.48}{output \per cycle} over \SI{0.25}{output \per cycle} and \SI{66}{\pico\joule \per output} over \SI{175}{\pico\joule \per output} respectively with 8-bit data. The significant difference in performance scaling further motivates the need for different architecture variants.
Physical constraints and the nature of the expected workload are the primary factors determining which architecture best suits a given scenario. Ultimately, the choice depends on both application-specific requirements and system performance targets set by the designer. The low power consumption, minimal area overhead, and high bitcell density of \caesar make it well-suited for ultra-low-power systems, where area and power are more critical than performance. Conversely, when performance and energy efficiency take precedence over area, the additional versatility and parallelism provided by \carus offer greater benefits. Additionally, \carus operates autonomously, freeing up the \cpu for other tasks.

The results in \cref{tab:carus-caesar-kernels} offer another important insight: the energy reduction of the \nmc-enhanced kernels is lower than their throughput gain. In other words, the system draws more power when executing the same workload on the \nmc macros compared to the \cpu baseline.
To better analyze power contributions in the system, a breakdown of the average power consumption during the execution of 8-bit and 32-bit 2D convolution kernels is shown in \cref{fig:power-breakdown}. The first observation is that memory accesses represent one of the largest contributions to system power consumption in all cases, supporting \ac{cim} as a viable solution to make edge computing more energy efficient. In the \cpu case, memory accesses consume approximately as much power as the \cpu itself. The \caesar case shows a similar overall power consumption, where memory accesses account for almost \SI{70}{\percent} of the total power, half of which is used to fetch the kernel micro-instructions and destination addresses from the system memory. Despite this, we have previously shown that \caesar can process significantly more data than the \cpu with the same power. The \cpu disadvantage is primarily due to the explicit instructions required to move the input data between the system memory and the \cpu \acp{gpr}, which are not necessary in the \nmc architectures. In addition, the \caesar power breakdown highlights the potential benefits of a dedicated controller that generates instructions on-the-fly instead of reading them from the system memory. This would allow \caesar to achieve the same level of autonomy as \carus and possibly a similar energy efficiency at the cost of reduced flexibility. Finally, \carus case confirms the advantages of exploiting a vector-capable architecture that makes the power contribution of its \cpu-based controller negligible compared the accesses to its four \vrf \sram banks (twice as \caesar's), which account for \SI{60}{\percent} of the total system power, and significantly more than its computing resources, that consume less power than the system \cpu despite producing far more output samples in the same amount of time.

\begin{figure}[htb]
    \centering
    \includesvg[width=\linewidth]{power-breakdown-conv2d.svg}
    \caption{Average power breakdown in \cpu (grey), \caesar (red), and \carus (cyan) for 8-bit and 32-bit 2D convolution. Post-layout simulation data (\SI{65}{\nano\meter}, $f_{CLK}=\SI{250}{\mega\hertz}$)}.
    \label{fig:power-breakdown}
    \vspace{-4mm}
\end{figure}

\subsubsection{Anomaly detection end-to-end application}
\noindent To better assess the performance of \caesar and \carus on a representative workload, the full \emph{Anomaly Detection} TinyML algorithm\cite{anomaly_det} is deployed on the \heeperator testbench system (\cref{fig:heeperator}) and compared against a multi-core \cpu cluster featuring the CV32E40P \riscv core\cite{riscy} with its \ac{dsp}-enhanced RV32IMCXcv \isa. The application implements an autoencoder composed of ten matrix-vector multiplication layers with ReLU activation functions.
\Cref{tab:tinyML_app} reports the cycle count, energy consumption, and area of a minimal system configured with a single-, dual-, and quad-core CV32E40P, as well as a tiny \riscv CV32E20 core (referred to as micro-riscy in \cite{riscv_comp}) coupled with either \caesar or \carus. The \cpu-based configurations include a single \SI{32}{\kibi\byte} \sram bank as the L1 data cache, which is replaced with the \nmc devices in the respective configurations. To ensure a fair comparison, the multi-core \cpu configurations are evaluated under the assumption of ideal linear scaling in both performance and area. Additionally, to focus the comparison on the computational capabilities of the systems, the contribution of the instruction memory is excluded from the energy consumption analysis.
As highlighted in the table, \caesar achieves a \SI{30}{\percent} higher throughput and \SI{20}{\percent} lower energy consumption compared to the single-core system, while also occupying \SI{10}{\percent} less area. However, it is slower and less energy-efficient than the multi-core configurations, making it an ideal candidate for area-constrained devices. On the other hand, \carus performs within \SI{13}{\percent} of the quad-core system while consuming half the energy and requiring less area than the dual-core configuration. This is a significant result, as it demonstrates the scalability potential of a multi-instance \carus-based data memory subsystem for energy-constrained applications.

\begin{table}[htb]
    \centering
    \caption{Execution time and energy improvements and area overhead of \caesar, \carus, and two multi-core CPU-based systems (dual- and quad-core) with respect to a single-core CPU-based system when running the \emph{Anomaly Detection} TinyML application.}
    \label{tab:tinyML_app}
    \begin{adjustbox}{width=1\linewidth}
    \begin{threeparttable}
    \begin{tabular}{l|c|cccc}
        \toprule
        \begin{tabular}[c]{@{}c@{}}\textbf{Metric}\end{tabular} & \begin{tabular}[c]{@{}c@{}}\textbf{CV32E40P}\tnote{a}\\(1 core)\end{tabular}  & \begin{tabular}[c]{@{}c@{}}\textbf{CV32E40P}\tnote{a}\\(2 cores)\end{tabular} & \begin{tabular}[c]{@{}c@{}}\textbf{CV32E40P}\tnote{a}\\(4 cores)\end{tabular} & \begin{tabular}[c]{@{}c@{}}\textbf{\caesar}\\\textbf{+ CV32E20}\tnote{b}\end{tabular} &\begin{tabular}[c]{@{}c@{}}\textbf{\carus}\\\textbf{+ CV32E20}\tnote{b}\end{tabular} \\ \midrule

        \begin{tabular}{@{}l@{}}Cycle count \end{tabular}   & \num{561e3} &  \textdownarrow~\SI{2.00}{\times} & \textbf{\textdownarrow~\SI{4.00}{\times}} & \textdownarrow~\SI{1.29}{\times} & \textdownarrow~\SI{3.55}{\times} \\ \midrule
        \begin{tabular}{@{}l@{}}Energy\tnotemid{c}\\\relax [\SI{}{\micro\joule}] \end{tabular}   & \num{13.5} & \textdownarrow~\SI{1.37}{\times} & \textdownarrow~\SI{1.67}{\times} & \textdownarrow~\SI{1.20}{\times} & \textbf{\textdownarrow~\SI{2.36}{\times}} \\ \midrule
        \begin{tabular}{@{}l@{}}Post-layout\\area [\SI{}{\micro\meter\squared}]\end{tabular} & \num{350e3} & \textuparrow~\SI{1.43}{\times} & \textuparrow~\SI{2.29}{\times} & \textbf{\textdownarrow~\SI{0.90}{\times}} & \textuparrow~\SI{1.36}{\times} \\ \bottomrule
    \end{tabular}
    \begin{tablenotes}[]
        \item[a] With RV32IMCXcv \isa.
        \item[b] With RV32E \isa.
        \item[c] At \SI{250}{\mega\hertz}; post-layout data.
    \end{tablenotes}
    \end{threeparttable}
    \end{adjustbox}
\end{table}


\subsection{Comparison with the State of the Art}
\label{sec:soa-comparison}

\noindent To better position the proposed \caesar and \carus \nmc architectures in the context of state-of-the-art \ac{cim} solutions, we compare them against recent representative\imc and \nmc \acp{ic}:
\begin{enumerate*}[label=(\arabic*)]
    \item BLADE \cite{blade-tc}, an in-\sram computing architecture that utilizes local wordline groups to perform energy-efficient computations,
    \item Computational SRAM (C-SRAM)\cite{CSRAM}, a hybrid \imc and \nmc scalable integrated vector unit, and
    \item Vecim \cite{wang2024vecim}, a high-performance vector coprocessor based on \cite{cavalcante2020ara} that exploits \imc to offer top-in-class energy efficiency.
\end{enumerate*}
\Cref{tab:soa-comparison} provides a quantitative and qualitative overview of the features, architecture, physical characteristics, and performance of these designs compared to \caesar and \carus despite their technology, size and architecture differences.
To establish a fair comparison from a terchnological standpoint among the different works, BLADE's and C-SRAM's physical characteristics were normalized by applying approximate scaling factors based on the scaling of an \sram bitcell from their original technologies to the target \SI{65}{\nano\meter} CMOS technology at the basis on \caesar, \carus, and Vecim. 
Since both designs are memory-dominated, the scaling factors were applied to both the memory and logic area, thus resulting in optimistic \SI{65}{\nano\meter} values for BLADE and C-SRAM. The energy scaling factors have been defined as the ratio between the energy cost of a read operation from a 6T or 8T \sram of equivalent array size implemented in a \SI{65}{\nano\meter} technology and the original \SI{28}{\nano\meter} (BLADE) or \SI{22}{\nano\meter} (C-SRAM) nodes.

Among the considered architectures, \carus achieves the highest peak energy efficiency, that is \SI{6}{\percent} higher than Vecim and \SI{21}{\percent} higher than the multi-array BLADE implementation. \caesar excels in memory density while maintaining high energy efficiency, especially if coupled with a dedicated controller mentioned previously. The Vecim vector coprocessor delivers the best peak performance thanks to its highly parallel execution units while offering a throughput per area within \SI{15}{\percent} to the multi-array BLADE instance. A similar performance density is expected from \carus instances with a higher lane count, since its throughput scales almost linearly with the number of \acp{alu}, while the area overhead of the additional logic and the smaller \vrf banks is contained.

\begin{table*}[bt]
    \centering
    \caption{Comparison with existing state-of-the-art \imc and \nmc solutions.}
    \label{tab:soa-comparison}
    \begin{adjustbox}{width=1\linewidth}
    \begin{threeparttable}
    \begin{tabular}{l|cc|cc|c|c|c|}
        \toprule
        {} & \multicolumn{2}{c|}{\begin{tabular}[c]{@{}c@{}}\textbf{BLADE}\cite{blade-tc}\end{tabular}} & \multicolumn{2}{c|}{\begin{tabular}[c]{@{}c@{}}\textbf{C-SRAM}\cite{CSRAM,CSRAM-2}\end{tabular}} & \begin{tabular}[c]{@{}c@{}}\textbf{Vecim}\cite{wang2024vecim}\end{tabular} & \begin{tabular}[c]{@{}c@{}}\textbf{\caesar} (this work)\end{tabular} & \begin{tabular}[c]{@{}c@{}}\textbf{\carus} (this work)\end{tabular} \\ \midrule[0.2pt]
        
        \acs{cim} type & \multicolumn{2}{c|}{\acs{imc}} & \multicolumn{2}{c|}{\acs{imc} + \acs{nmc}} & \acs{imc} + \acs{nmc} & \acs{nmc} & \acs{nmc} \\ \midrule[0.2pt]
        
        Array instances & \multicolumn{2}{c|}{16 \texttimes\xspace \SI{2}{\kibi\byte}} & \multicolumn{2}{c|}{4 \texttimes\xspace \SI{8}{\kibi\byte}} & 1 \texttimes\xspace \SI{16}{\kibi\byte} (4 lanes) & 1 \texttimes\xspace \SI{32}{\kibi\byte} & 1 \texttimes\xspace \SI{32}{\kibi\byte} (4 lanes) \\ \midrule[0.2pt]
        
        \acs{sram} type & \multicolumn{2}{c|}{Foundry 6T, custom array} & \multicolumn{2}{c|}{Foundry 8T, custom array} & Foundry 8T, custom array & Foundry 6T & Foundry 6T \\ \midrule[0.2pt]

        Usefull bitcell density\tnotemid{a} [\SI{}{\percent}] & \multicolumn{2}{c|}{\textbf{\num{53.5}}} & \multicolumn{2}{c|}{\num{20.3}} & 1.7 & \textbf{\num{54.0}} & \num{33.0} \\ \midrule[0.2pt]
        
        Deployment constraints & \multicolumn{2}{c|}{\begin{tabular}[c]{@{}c@{}}\tabitem Word alignment\\ \tabitem Data placement (LG)\end{tabular}} & \multicolumn{2}{c|}{\begin{tabular}[c]{@{}c@{}}\tabitem Word alignment\\ \tabitem Data replication ($D, \overline{D}$)\end{tabular}} & \begin{tabular}[c]{@{}c@{}}\tabitem Vector alignment\end{tabular} & \tabitem Word alignment & \begin{tabular}[c]{@{}c@{}}\tabitem Vector alignment\end{tabular} \\ \midrule[0.2pt]
        
        Instance \ac{alu} features & \multicolumn{2}{c|}{\begin{tabular}[c]{@{}c@{}}32-bit \acs{simd} \ac{alu}\end{tabular}} & \multicolumn{2}{c|}{\begin{tabular}[c]{@{}c@{}}128-bit \acs{simd} \ac{alu}\end{tabular}} & 4 \texttimes\xspace 256-bit \acs{simd} \acp{alu} & \begin{tabular}[c]{@{}c@{}}2-cycle 32-bit \acs{simd} \ac{alu}\end{tabular} & \begin{tabular}[c]{@{}c@{}}4 \texttimes\xspace 32-bit \acs{simd} \acp{alu}\end{tabular} \\ \midrule[0.2pt]

        Multiplier architecture & \multicolumn{2}{c|}{1-bit Add and shift} & \multicolumn{2}{c|}{Add and shift} & Double-rate bit-parall. & \begin{tabular}[c]{@{}c@{}}4 \texttimes\xspace 17-bit multipliers\end{tabular} & \begin{tabular}[c]{@{}c@{}}1 \texttimes\xspace 16-bit mult. per \acs{alu}\end{tabular} \\ \midrule[0.2pt]

        Supported operations & \multicolumn{2}{c|}{\begin{tabular}[c]{@{}c@{}}\tabitem ADD/SUB\\ \tabitem MULT \tabitem Logic\end{tabular}} & \multicolumn{2}{c|}{\begin{tabular}[c]{@{}c@{}}\tabitem ADD/SUB \tabitem MULT\\ \tabitem \acs{mac} \tabitem Logic \tabitem Copy\\ \tabitem Shuffle \tabitem Comparison\\ \tabitem Logical \& arith. shift\end{tabular}} & \begin{tabular}[c]{@{}c@{}}\tabitem \rvv instructions\\(integer and\\floating-point)\end{tabular} & \begin{tabular}[c]{@{}c@{}}\tabitem ADD/SUB \tabitem MULT\\ \tabitem \acs{mac} \tabitem Logical shift\\ \tabitem Logic \tabitem Comparison\end{tabular} & \begin{tabular}[c]{@{}c@{}}\tabitem RV32EC \isa\\ \tabitem ADD/SUB \tabitem MULT\\ \tabitem \acs{mac} \tabitem Comparison\\ \tabitem Logical \& arith. shift\\\tabitem Logic \tabitem Copy \tabitem Slide\end{tabular} \\ \midrule[0.2pt]
        
        Technology & \SI{28}{\nano\meter} & \SI{65}{\nano\meter}\tnote{b} & \SI{22}{\nano\meter} & \SI{65}{\nano\meter}\tnote{b} & \SI{65}{\nano\meter} & \SI{65}{\nano\meter} & \SI{65}{\nano\meter} \\ \midrule[0.2pt]
        
        Area [\SI{}{\micro\meter\squared}] & \num{64e3} & \num{580e3} & \num{17.5e3} & N/A\tnote{c} & \num{4e6} & \num{256e3} & \num{415e3} \\ \midrule[0.2pt]

        Nominal freq. [\SI{}{\mega\hertz}] & \num{2200} & \num{330}\tnote{d} & \num{1000} & \num{330}\tnote{d} & \num{250} & \num{330} & \num{330} \\ \midrule[0.2pt]
    
        
        \begin{tabular}[c]{@{}c@{}}Peak throughput\tnotemid{e} [\SI{}{\giga OPS}]\end{tabular} & \textbf{\num{35.2}} & \num{5.3} & \num{10.7} & \num{3.5} & \textbf{\num{31.8}} & \num{1.32} & \num{2.64} \\ \midrule[0.2pt]

        Energy eff. [\SI{}{\giga OPS\per\watt}] & \num{830.7}\tnote{f} & \num{254.2}\tnote{f} & \num{52.0} & \num{13.2} & \num{289.1} & \num{200.3} (\num{421.9}\tnote{f}~)\tnote{f} & \textbf{\num{306.7}}\tnote{g} \\ \midrule[0.2pt]

        Area eff. [\SI{}{\giga OPS\per\milli\meter\squared}] & \num{550}\tnote{f} & \textbf{\num{9.1}} & \num{611} & N/A\tnote{c} & \num{8.0} & \num{5.2} & \num{6.4} \\ \bottomrule
    \end{tabular}
    \footnotesize
    \begin{tablenotes}[]
        \item[a] Area of useful bitcells (i.e., replicated data is not considered), normalized to 6T bitcell area.
        \item[b] Area, timing, and power at \SI{65}{\nano\meter} were computed from \SI{28}{\nano\meter} ones based on commercial \sram scaling on the same technologies. This is a conservative, best-case scaling factor for the standard-cell parts of the design.
        \item[c] Area estimation from \SI{22}{\nano\meter} data is not trivial because of C-SRAM's mixed \imc/\nmc design, so it's not provided.
        \item[d] Assumed to match the operating frequency of the \SI{65}{\nano\meter} \SI{32}{\kibi\byte} \sram used as reference to define \caesar and \carus timing constraints.
        \item[e] 8-bit MAC operations unless specified otherwise. One \acs{mac} operation is considered as two elementary operations (one multiplication and one addition).
        \item[f] Without controller.
        \item[g] Average power from post-layout simulation while running a matrix multiplication kernel.
    \end{tablenotes}
    \end{threeparttable}
    \end{adjustbox}
\end{table*}

To better compare the performance of the considered \ac{cim} architectures, \Cref{tab:soa-performance} reports the peak throughput and the energy consumption of BLADE, C-SRAM, \caesar, and \carus when executing a matrix multiplication on different data widths. The throughput of BLADE and C-SRAM was estimated based on the values reported in the respective articles for multiplication operations. Because they have been implemented as \SI{2}{\kibi\byte} and \SI{4}{\kibi\byte} macros, we considered 16 and 8 instances, respectively, to match the \SI{32}{\kibi\byte} memory capacity of \caesar and \carus. A single \SI{32}{\kibi\byte} instance of BLADE is also included to show its scaling versus capacity. Structural hazards and data replication for data sizes exceeding the array's dimension are not considered for BLADE and C-SRAM. Similarly, we only consider the proportional increase in static leakage power for a \SI{32}{\kibi\byte} BLADE subarray when evaluating its energy consumption, not taking into account the increase in dynamic power.
All these choices were made not to penalize or underestimate the performance of BLADE and C-SRAM, placing them under optimal conditions in an attempt to provide a fair comparison with the \nmc architectures proposed in this work despite the different technology nodes.
\carus achieves \SI{1.31}{\times} and \SI{1.97}{\times} faster execution time than the multi-array BLADE implementation on 16-bit and 8-bit data respectively, performing \SI{2.1}{\times} worse only in the 8-bit kernel because of the better scaling of BLADE's add-and-shift multiplier architecture for smaller data types. However, BLADE's performance may be degraded in real-world scenarios because of unsupported inter-bank operations. \carus is also the most energy-efficient design, consuming up to \SI{3}{\times} less energy than BLADE on 32-bit data. \caesar, on the other hand, achieves the same 32-bit performance of BLADE at \SI{60}{\percent} its power consumption and half its area. The single-array BLADE instance and the multi-array C-SRAM perform worse than \carus or the multi-array BLADE in all cases. In this regard, it must be noted that the C-SRAM data has been extracted from silicon measurements instead of post-layout simulations, possibly explaining the discrepancy.

With the increasing deployment of computationally intensive algorithms on low-power devices, reduced-precision models utilizing sub-byte data types are becoming increasingly popular. \ac{cim} architectures that employ serial multipliers, such as the \emph{add-and-shift} units in BLADE and C-SRAM, would achieve higher energy efficiency than the proposed \nmc devices when dealing with such models. However, as discussed in \cref{sec:arch}, the \acp{alu} of \caesar and \carus were specifically designed for standard data types,which gives no advantage when handling lower-precision models, as their data would be sign-extended to 8 bits.


\begin{table*}[hbt]
\caption{Peak performance comparison on matrix multiplications.}
\label{tab:soa-performance}
\centering
    \begin{threeparttable}
    \scriptsize
    \begin{tabular}{ll|r|r|r|r|r|r|r|r|}
    \toprule
    \multicolumn{2}{c|}{\multirow{2}{*}{Metric and bitwidth}} & \multicolumn{2}{c|}{\begin{tabular}[c]{@{}c@{}}\textbf{BLADE}\tnote{a,b}\\16 \texttimes\xspace \SI{2}{\kibi\byte}\end{tabular}} & \multicolumn{2}{c|}{\begin{tabular}[c]{@{}c@{}}\textbf{BLADE}\\ 1 \texttimes\xspace \SI{32}{\kibi\byte}\end{tabular}} & \multicolumn{2}{c|}{\begin{tabular}[c]{@{}c@{}}\textbf{C-SRAM}\tnote{a,b}\\ 8 \texttimes \SI{4}{\kibi\byte}\end{tabular}} & \multicolumn{1}{c|}{\begin{tabular}[c]{@{}c@{}}\textbf{\caesar}\\ (this work)\end{tabular}} & \begin{tabular}[c]{@{}c@{}}\textbf{\carus}\\ (this work)\end{tabular} \\

    & & \multicolumn{1}{c|}{{\SI{28}{\nano\meter}}} & \multicolumn{1}{c|}{{\SI{65}{\nano\meter}}} & \multicolumn{1}{c|}{{\SI{28}{\nano\meter}}} & \multicolumn{1}{c|}{{\SI{65}{\nano\meter}}} & \multicolumn{1}{c|}{{\SI{22}{\nano\meter}}} & \multicolumn{1}{c|}{{\SI{65}{\nano\meter}}} & \multicolumn{1}{c|}{{\SI{65}{\nano\meter}}} & \multicolumn{1}{c|}{{\SI{65}{\nano\meter}}} \\ \midrule[0.5pt]

    \multirow{3}{*}{Cycle count} & 8-bit\tnote{d} & \multicolumn{2}{c|}{\textbf{\num{12.8e3}}} & \multicolumn{2}{c|}{\num{204.8e3}} & \multicolumn{2}{c|}{\num{19.2e3}} & {\num{51.2e3}} & \num{26.6e3} \\
    
    & 16-bit\tnote{e} & \multicolumn{2}{c|}{\num{25.6e3}} & \multicolumn{2}{c|}{\num{409.6e3}} & \multicolumn{2}{c|}{\num{38.4e3}} & {\num{51.2e3}} & \textbf{\num{19.5e3}} \\
    
    & 32-bit\tnote{f} & \multicolumn{2}{c|}{\num{51.2e3}} & \multicolumn{2}{c|}{\num{819.2e3}} & \multicolumn{2}{c|}{\num{76.8e3}} & {\num{51.2e3}} & \textbf{\num{26.0e3}} \\ \midrule[0.2pt]
    
    \multirow{3}{*}{\begin{tabular}[l]{@{}l@{}}Execution time\tnote{c}\\\relax [\SI{}{\micro\second}]\end{tabular}} & 8-bit\tnote{d} & {5.8} & {\textbf{38.8}} & {93} & {620} & {19.3} & {58.1} & {155} & 80.6 \\
    
    & 16-bit\tnote{e} & {11.6} & {77.5} & {186} & {1240} & {38.4} & {116} & {155} & \textbf{59.1} \\
    
    & 32-bit\tnote{f} & {23.3} & {155} & {372} & {2480} & {76.8} & {232} & {155} & \textbf{78.8} \\ \midrule[0.2pt]
    
    \multirow{3}{*}{\begin{tabular}[l]{@{}l@{}}Energy\\\relax [\SI{}{\pico\joule\per MAC}]\end{tabular}} & 8-bit\tnote{d} & {2.4} & {7.9} & {13} & {43} & {38.8} & {150} & {16.3} & \textbf{6.8} \\
    
    & 16-bit\tnote{e} & {8.1} & {26.7} & {29.4} & {97.1} & {155} & {600} & {32} & \textbf{12.0} \\
    
    & 32-bit\tnote{f} & {31.1} & {103} & {96.9} & {320} & {621} & {2400} & {61.8} & \textbf{31.2} \\ \bottomrule
    \end{tabular}
    \footnotesize
    \begin{tablenotes}[]
        \item [a] Data replication latency is neglected (best-case scenario).
        \item [b] Inter-bank data placement hazards (e.g., \acs{mac} source operands in different banks) are neglected (best-case scenario).
        \item [c] At nominal frequency.
        \item [d] $A[10,10] \times B[10,1024]$
        \item [e] $A[10,10] \times B[10,512]$
        \item [f] $A[10,10] \times B[10,256]$
    \end{tablenotes}
    \end{threeparttable}
    \vspace{-2mm}
\end{table*}

%% file: conclusion.tex
\section{Conclusion}
\label{sec:conclusion}

\noindent This article has presented \caesar and \carus, two novel architectural variations of a new \nmc approach that targets energy-efficient and low-power \ac{soc} designs for edge computing devices running TinyML algorithms on the edge. 
Leveraging the \nmc paradigm, they enable the execution of arithmetic operations next to the memory array without the need to transfer data through the system bus.
By performing computation on vectors in a \ac{simd} manner, they can improve the throughput and energy performance of conventional architecture relying on energy-efficient \acp{cpu} or state-of-the-art \acs{imc}/\ac{nmc} architectures.
Experimental results have shown that \caesar and \carus achieve a timing speed-up of up to \SI{25.8}{\times} and \SI{50.0}{\times}, as well as a reduction in energy of \SI{23.2}{\times} and \SI{33.1}{\times} compared to a scalar architecture based on \riscv in a matrix multiplication kernel. In particular, a \SI{65}{\nano\meter} implementation of \carus demonstrates a peak energy efficiency of \SI{306.7}{\giga OPS \per \watt} with 8-bit data. In addition, their software-friendly digital-based design approach will ease their integration into existing systems without any major changes for optimal performance across different technology nodes. 

\section{Acknowledgements}
This work was supported in part by the Swiss State Secretariat for Education, Research, and Innovation (SERI) through the SwissChips research project.

%% file: acronyms.tex

\begin{acronym}
    \acro{ai}[AI]{Artificial Intelligence}
    \acro{alu}[ALU]{Arithmetic Logic Unit}
    \acro{ann}[ANN]{Artificial Neural Network}
    \acro{api}[API]{Application Programming Interface}
    \acro{asic}[ASIC]{Application-Specific Integrated Circuit}
    \acro{cgra}[CGRA]{Coarse-Grained Reconfigurable Architecture}
    \acro{cim}[CIM]{Compute-In-Memory}
    \acro{cnn}[CNN]{Convolutional Neural Network}
    \acro{cpu}[CPU]{Central Processing Unit}
    \acro{csr}[CSR]{Control and Status Register}
    \acro{dma}[DMA]{Direct Memory Access}
    \acro{dnn}[DNN]{Deep Neural Network}
    \acro{dram}[DRAM]{Dynamic Random Access Memory}
    \acroplural{dram}[DRAMs]{Dynamic Random Access Memories}
    \acro{dsp}[DSP]{Digital Signal Processor}
    \acro{ecpu}[eCPU]{embedded CPU}
    \acro{emem}[eMEM]{embedded Memory}
    \acro{envm}[ENMV]{Embedded Non-Volatile Memory}
    \acroplural{envm}[ENMVs]{Embedded Non-Volatile Memories}
    \acro{fft}[FFT]{Fast Fourier Transform}
    \acro{fll}[FLL]{Frequency-Locked Loop}
    \acro{fpga}[FPGA]{Field-Programmable Gate Array}
    \acro{fsm}[FSM]{Finite State Machine}
    \acro{gcc}[GCC]{GNU Compiler Collection}
    \acro{gemm}[GEMM]{General Matrix Multiplication}
    \acro{gpr}[GPR]{General-Purpose Register}
    \acro{hdl}[HDL]{Hardware Description Language}
    \acro{ic}[IC]{Integrated Circuit}
    \acro{imc}[IMC]{In-Memory Computing}
    \acro{iot}[IoT]{Internet of Things}
    \acro{isa}[ISA]{Instruction Set Architecture}
    \acro{mac}[MAC]{Multiply-and-Accumulate}
    \acro{mcu}[MCU]{Microcontroller Unit}
    \acro{ml}[ML]{Machine Learning}
    \acro{mram}[MRAM]{Magnetoresistive Random Access Memory}
    \acroplural{mram}[MRAMs]{Magnetoresistive Random Access Memories}
    \acro{nmc}[NMC]{Near-Memory Computing}
    \acro{pcm}[PCM]{Phase-Change Memory}
    \acroplural{pcm}[PCMs]{Phase-Change Memories}
    \acro{pe}[PE]{Processing Element}
    \acro{pim}[PIM]{Processing-In-Memory}
    \acro{pvt}[PVT]{Process, Voltage, and Temperature}
    \acro{relu}[ReLU]{Rectified Linear Unit}
    \acro{risc}[RISC]{Reduced Instruction Set Computer}
    \acro{rom}[ROM]{Read-Only Memory}
    \acroplural{rom}[ROMs]{Read-Only Memories}
    \acro{rram}[RRAM]{Resistive Random Access Memory}
    \acroplural{rram}[RRAMs]{Resistive Random Access Memories}
    \acro{rtl}[RTL]{Register Transfer Level}
    \acro{rvv}[RVV]{RISC-V Vector Extension}
    \acro{sew}[SEW]{Selected Element Width}
    \acro{simd}[SIMD]{Single Instruction Multiple Data}
    \acro{soc}[SoC]{System-on-Chip}
    \acroplural{soc}[SoCs]{Systems-on-Chip}
    \acro{sram}[SRAM]{Static Random-Access Memory}
    \acroplural{sram}[SRAMs]{Static Random-Access Memories}
    \acro{vcd}[VCD]{Value Change Dump}
    \acro{vpu}[VPU]{Vector Processing Unit}
    \acro{vrf}[VRF]{Vector Register File}
    \acro{x-heep}[X-HEEP]{eXtendable Heterogeneous Energy-efficient Platform}
\end{acronym}

%% file: authors.tex
\begin{IEEEbiography}[{\includegraphics[width=1in,height=1.25in,clip,keepaspectratio]{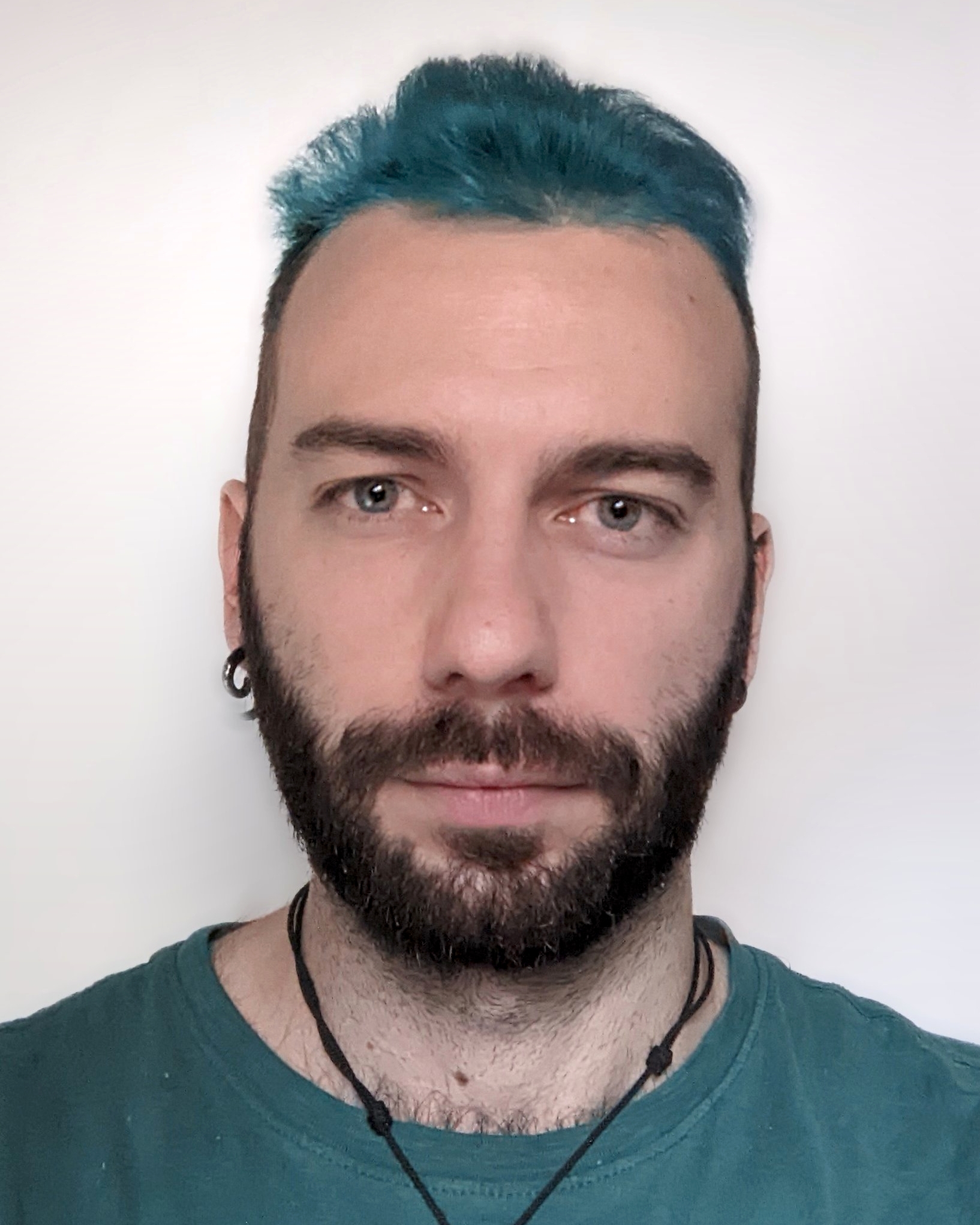}}]{Michele Caon}
    is a postdoctoral researcher at the Embedded Systems Laboratory of EPFL. He received his B.S. and M.S., and Ph.D. at the Electronics and Telecommunications Department at Politecnico di Torino in 2017, 2019, and 2024. His research interests are innovative digital, integrated, programmable computing circuits and systems. He is currently working on near-memory computing circuits embedded in heterogeneous systems-on-chip for edge computing applications.
\end{IEEEbiography}

\begin{IEEEbiography}[{\includegraphics[width=1in,height=1.25in,clip,keepaspectratio]{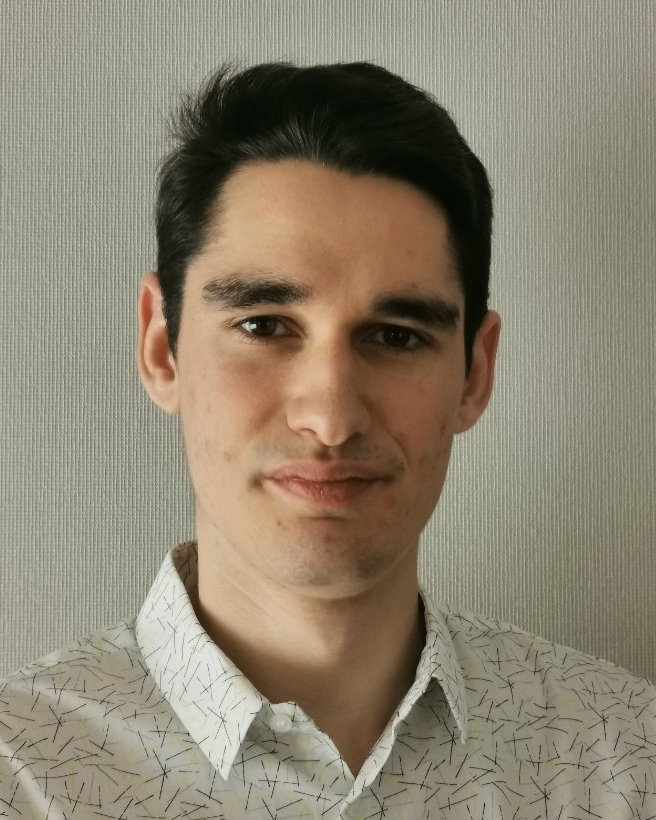}}]{Clément Choné}
    received the MSc degree in Micro and Nanotechnologies for integrated systems from Grenoble-INP Phelma, in 2022. He is currently working towards his PhD degree with the Embedded Systems Laboratory (ESL) of EPFL, Switzerland, under the supervision of Prof. David Atienza and Prof. Andreas Burg. His main research interests include the development of near-memory computing accelerators and low-power logic and memory systems using innovative circuit styles to suppress leakage current in low-power systems.
\end{IEEEbiography}

\begin{IEEEbiography}[{\includegraphics[width=1in,height=1.25in,clip,keepaspectratio]{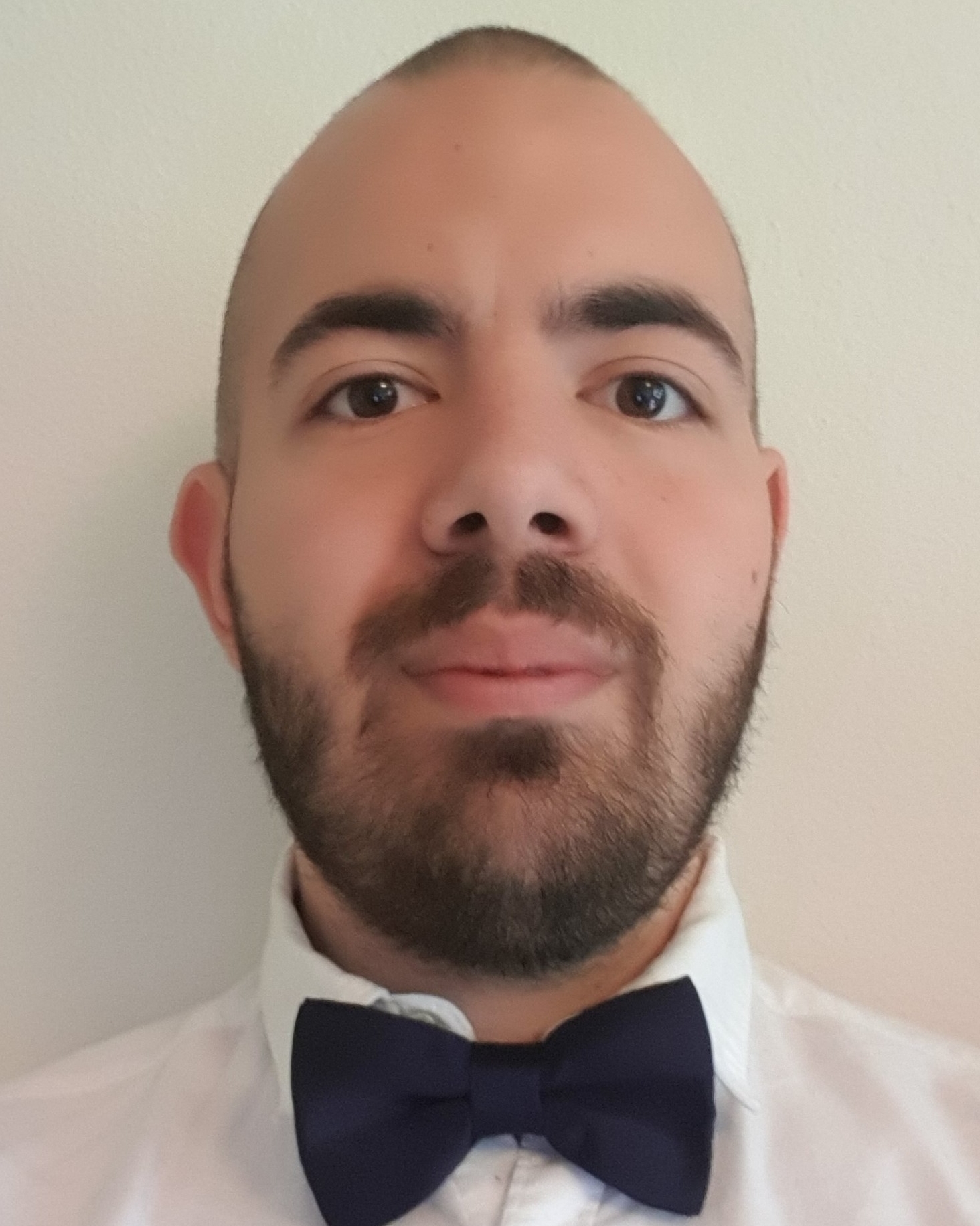}}]{Pasquale Davide Schiavone}
    is a postdoctoral researcher at the Embedded Systems Laboratory of EPFL and Director of Engineering of the OpenHW Group. He obtained the Ph.D. title at the Integrated Systems Laboratory of ETH Zurich in the Digital Systems group in 2020 and the BSc. and MSc. from "Politecnico di Torino" in computer engineering in 2013 and 2016, respectively. His main activities are the \riscv \cpu design and low-power energy-efficient computer architectures for smart embedded systems and edge-computing devices.
\end{IEEEbiography}

\begin{IEEEbiography}[{\includegraphics[width=1in,height=1.25in,clip,keepaspectratio]{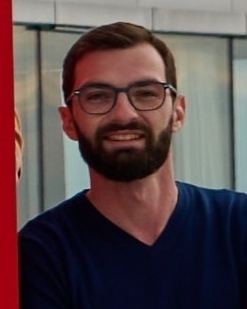}}]{Alexandre Levisse}
    received his Ph.D. degree in Electrical Engineering from CEA-LETI and Aix-Marseille University, France, in 2017. From 2018 to 2021, he was a postdoctoral researcher at the Embedded Systems Laboratory of EPFL. From 2021, he works as a scientist in EPFL. His research interests include circuits and architectures for emerging memory and transistor technologies as well as in-memory computing and accelerators.
\end{IEEEbiography}

\begin{IEEEbiography}[{\includegraphics[width=1in,height=1.25in,clip,keepaspectratio]{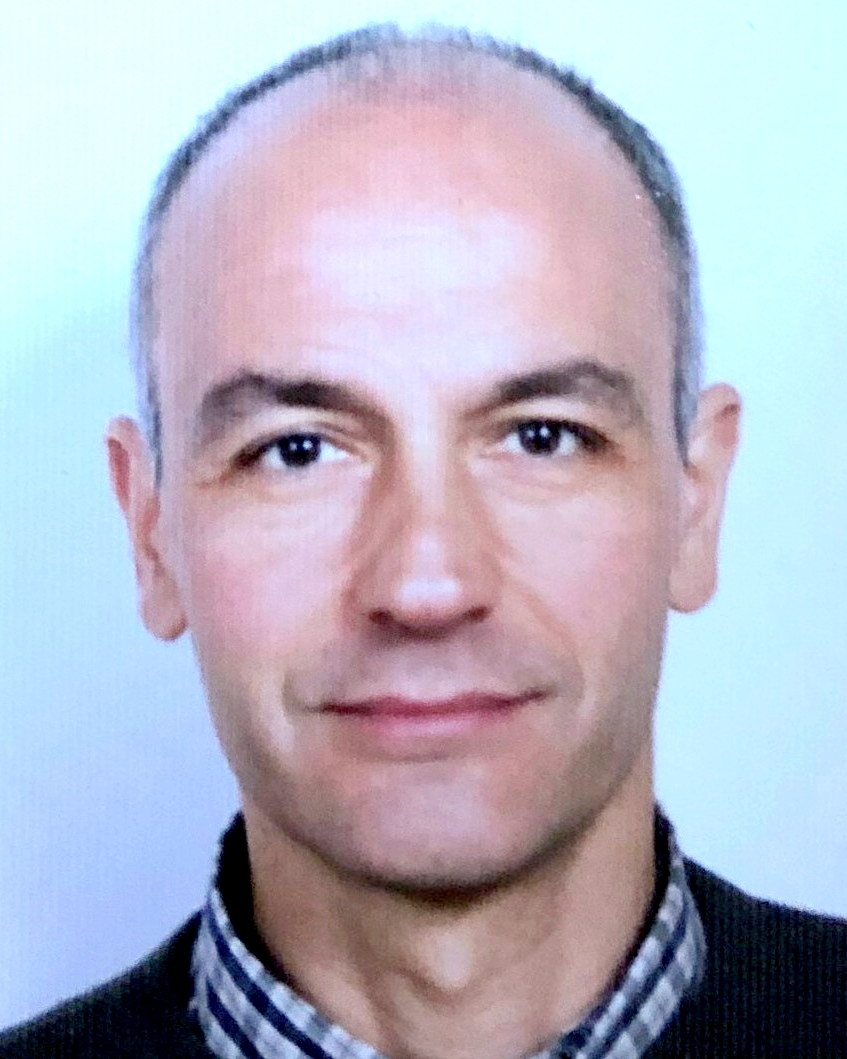}}]{Guido Masera}
    (SM'07)   received   the   Dr.-Ing. (summa cum laude) and Ph.D. degrees in Electronic Engineering from Politecnico di Torino, Italy. He has been a professor with the Electronics and Telecommunications Department,  Politecnico di Torino, since 1992. His research interests focus digital integrated circuits and systems, with a special emphasis on high-performance architectures for communications, forward error correction, image and video coding, cryptography and hardware accelerators for machine learning. He has more than 200 publications, two patents and was a designer of several ASIC components. Dr. Masera is an Associate  Editor of MDPI Electronics and a former Associate Editor of the IEEE Transactions on Circuits and Systems I, IEEE Transactions on Circuits and Systems II and the IET Circuits, Devices \& Systems.
\end{IEEEbiography}

\begin{IEEEbiography}[{\includegraphics[width=1in,height=1.25in,clip,keepaspectratio]{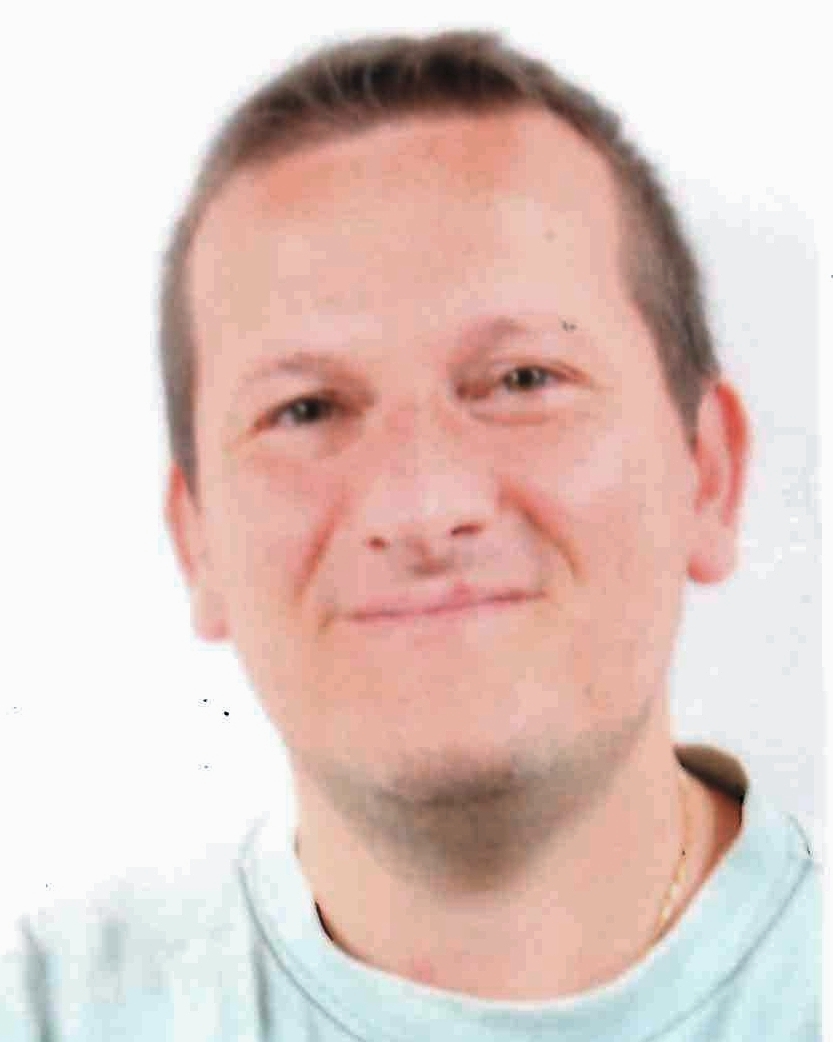}}]{Maurizio Martina}
    received the Dr.-Ing. and Ph.D. degrees in electronic engineering and electronic and communications engineering from Politecnico di Torino, Italy, in 2000 and 2004. He is a Professor with the Electronics and Telecommunications Department, Politecnico di Torino, since 2014. His research interests include computer architecture and VLSI design of digital integrated circuits for image and video coding, forward error correction, cryptography and artificial intelligence. He has more than 100 publications and holds two patents. 
    He served as an Associate Editor of the IEEE Transactions on Circuits and Systems—I and as a Guest Editor of several special issues, including BioCAS 2017 special issue in IEEE Transactions of Biomedical Circuits and Systems and ISCAS 2023 special issue in IEEE Transactions on Circuits and Systems-II. He has been part of the organizing and technical committee of several IEEE conferences, including BioCAS 2017, AICAS 2020, and PRIME 2025.
\end{IEEEbiography}

\begin{IEEEbiography}[{\includegraphics[width=1in,height=1.25in,clip,keepaspectratio]{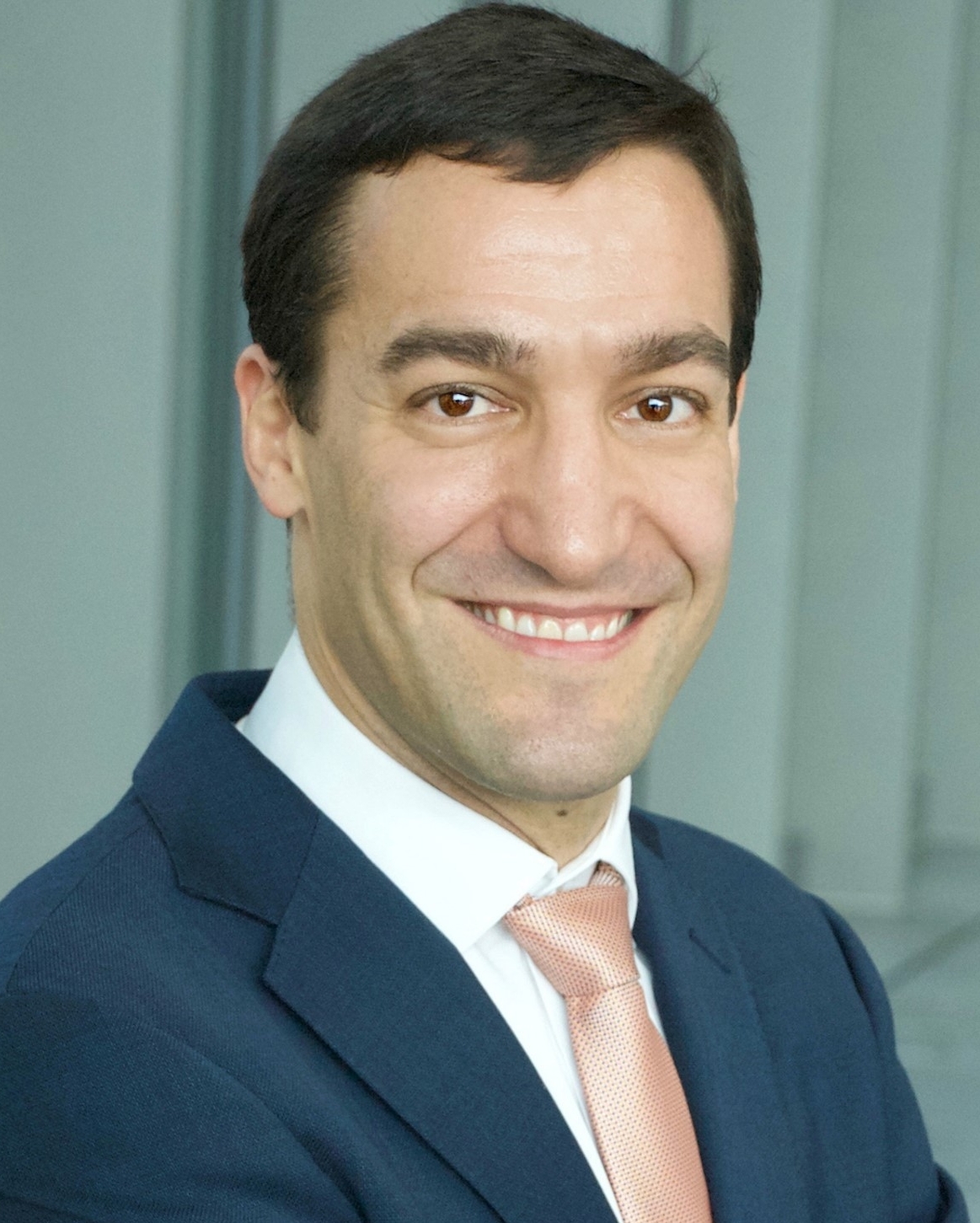}}]{David Atienza}
    (M'05-SM'13-F'16) received his MSc and PhD degrees in Computer Science and Engineering from Complutense Univ. of Madrid, Spain, and IMEC, Belgium, in 2001 and 2005. He is a professor of Electrical and Computer Engineering, Associate Vice President of Research Centers and Platforms, and heads the Embedded Systems Laboratory (ESL) at EPFL, Switzerland. His research interests include system-level design methodologies for high-performance multi-processor system-on-chip and low-power Internet-of-Things systems, including new thermal-aware design for 2-D/3-D MPSoCs and many-core servers, ultra-low power system architectures for wearable systems and edge AI computing, and memory hierarchy optimizations. He is a co-author of more than 450 publications and 14 patents. He serves as Editor-in-Chief of IEEE TCAD (period 2022-2025) and ACM CSUR (since 2024). Dr. Atienza received the 2024 Test-of-Time Best Paper Award at CODES+ISSS, the ICCAD 10-Year Retrospective Most Influential Paper Award, and the DAC Under-40 Innovators Award, among others. He is a Fellow of the ACM.
\end{IEEEbiography}